\documentclass{aa} 
\usepackage{graphicx}
\usepackage[normalem]{ulem} 
\usepackage{threeparttable}
\usepackage{epstopdf} 
\usepackage[T1]{fontenc}
\usepackage{hyphenat}

\usepackage{txfonts}
\begin{document}

   \title{Accretion of Uranus and Neptune from inward-migrating planetary embryos blocked by Jupiter and Saturn }
   \titlerunning{Accretion of Uranus and Neptune}

   \author{André Izidoro
          \inst{1,2,3}\fnmsep\thanks{izidoro.costa@gmail.com},
           Alessandro Morbidelli\inst{2},
           Sean N. Raymond\inst{4},
           Franck Hersant \inst{4},
           \and
           Arnaud Pierens\inst{4}
          }
\authorrunning{Izidoro et al.}
   \institute{Université de Bordeaux, Laboratoire d'Astrophysique de Bordeaux, UMR 5804, F-33270, Floirac, France \and
   University of Nice-Sophia Antipolis, CNRS, Observatoire de la Côte d’Azur, Laboratoire Lagrange, BP 4229, 06304 Nice Cedex 4, France
   \and Capes Foundation, Ministry of Education of Brazil, Brasília/DF 70040-020, Brazil.
         \and
             CNRS and Université de Bordeaux, Laboratoire d'Astrophysique de Bordeaux, UMR 5804, F-33270, Floirac, France\\
             }

   \date{Received ...; accepted...}

 \abstract{ Reproducing Uranus and Neptune remains a challenge for simulations of solar system formation. The ice giants' peculiar obliquities suggest that they both suffered giant collisions during their formation. Thus, there must have been an epoch of accretion dominated by collisions among large planetary embryos in the primordial outer solar system. We test this idea using N-body numerical simulations including the effects of a gaseous protoplanetary disk.  One strong constraint is that the masses of the ice giants are very similar -- the Neptune/Uranus mass ratio is $\sim1.18$.  We show that similar-size ice giants do indeed form by collisions between planetary embryos beyond Saturn. The fraction of successful simulations varies depending on the initial number of planetary embryos in the system, their individual and total masses. Similar-sized ice giants are consistently reproduced in simulations starting with 5-10 planetary embryos with initial masses of $\sim$3-6 ${\rm M_\oplus}$. We conclude that accretion from a population of planetary embryos is a plausible scenario for the origin of Uranus and Neptune.}


  \maketitle

%

\section{Introduction}

The formation of Uranus and Neptune is one of the longest-standing problems in solar system formation (Safronov, 1972; Levison \& Stewart, 2001; Thommes et al., 1999; 2002; Goldreich et al., 2004a,b; Morbidelli et al., 2012; Jakubik et al., 2012). The accretion timescale is strongly dependent on the amount of solid material available (i.e. the density of solids) and on the dynamical timescales (related to the orbital period) in the region of formation (e.g. Safronov, 1972). At their current positions, Uranus and Neptune's calculated accretion timescales are implausibly long (Levison and Stewart, 2001; Thommes et al., 2003) because of the low density in the protoplanetary disk (e.g. Weidschelling, 1977; Hayashi, 1981) and long dynamical timescales beyond $\sim$20 AU. Goldreich et al., (2004a,b) claimed to have solved the problem by assuming that Uranus and Neptune formed from a highly collisional disk of small particles, but Levison and Morbidelli (2007) later showed that the simulated evolution of the system is very different from what they envisioned analytically.

Studies of the formation and dynamical evolution of giant planets in our solar system (see a review by Morbidelli et al., 2012) as well as the discoveries of extrasolar hot-Jupiters (Cumming et al 2008; Mayor et al., 2011; Howard et al., 2012; Batalha et al., 2013; Fressin et al., 2013) and hot super-Earths (Mayor et al. 2009, 2011; Howard et al. 2010, 2012), have destroyed the belief that planets formed where they are now observed. Planetary migration seems to be a generic process of planetary formation. During the gas-disk phase planets exchange angular momentum  with their natal protoplanetary disk and migrate in a regime that depends on the planet mass (Ward, 1986; 1997). After gas disk dissipation, planet migration is also possible due to other mechanisms, such as tidal interaction of the planet with its host star (e.g., Rasio et al. 1996; Jackson et al. 2008),  gravitational scattering of planetesimals by the planet (eg. Fernandez and Ip, 1984; Hahn and Malhotra, 1999; Gomes, 2003) or mutual scattering between planets  (Thommes et al., 1999; Tsiganis et al., 2005; Nagasawa et al., 2008; Naoz et al., 2011).

The orbital structure of small body populations firmly supports the hypothesis that Saturn, Uranus and Neptune migrated outward after the gas-disk dispersed by interactions with a leftover disk of planetesimal (e.g. Fernandez \& Ip, 1984; Hahn \& Malhotra, 1999; Gomes, 2003). In the Nice Model  (Gomes et al., 2005; Morbidelli et al., 2005; 2007; Tsiganis et al., 2005; Levison et al., 2008; Levison et al., 2011; Nesvorny, 2011; Nesvorny \& Morbidelli 2012; Batygin et al., 2010; 2012) all giant planets  would have formed inside 15 AU. This may partially alleviate the accretion timescale problem. However, even in these more propitious conditions, the accretion of multiple $\sim$10 $M_\oplus$ planetary cores from planetesimals during the gas disk lifetime remains unlikely (Levison et al., 2010).

In fact, Levison et al (2010) were unable to repeatedly form giant planet cores by accretion of planetesimals via runaway (e.g., Wetherill \& Stewart 1989; Kokubo \& Ida 1996) and oligarch growth (e.g. Ida \& Makino 1993, Kokubo \& Ida, 1998; 2000). Planetary embryos and cores stir up neighboring planetesimals and increase their velocity dispersions.  Cores open gaps in the distribution of planetesimals around their orbits (Ida and Makino, 1993; Tanaka \& Ida, 1997) and this drastically reduces their rate of growth long before reaching masses comparable to those of the real ice giants' (Levison et al., 2010). 

A new model for the formation of planetary cores called pebble accretion may help solve this problem (Johansen, 2009; Lambrechts \& Johansen, 2012; Morbidelli and Nesvorny, 2012; Chambers, 2014; Kretke \& Levison, 2014).  In the pebble-accretion model, planetary cores grow from a population of seed planetesimals.  Planetesimals accrete pebbles spiralling towards the star due to gas drag (Johansen et al., 2009).  The formation of  multi-Earth-mass planetary cores can be extremely fast even in traditional disks such as the minimum mass solar nebula (Weidenschilling, 1977; Hayashi et al., 2011). The timescale accretion problem disappears even if Uranus and Neptune formed at their current locations (Lambrechts \& Johansen, 2014; Lambrechts et al., 2014).

It is unlikely, though, that Uranus and Neptune formed solely by pebble accretion. The ice giants have large obliquities (spin axis inclinations relative to their orbital planes): about 90 degrees for Uranus and about 30 degrees for Neptune. A planet accreting only small bodies should have a null obliquity (Dones and Tremaine, 1993; Johansen and Lacerda, 2010). Yet Jupiter is the only giant planet with a small obliquity. Saturn has a 26 degree obliquity but this is probably due to a spin-orbit resonance with Neptune (Ward and Hamilton, 2004; Hamilton and Ward, 2004; Boue et al., 2009). The terrestrial planets have a quasi-random obliquity distribution due to the giant impacts that suffered during their formation (Agnor et al 1999; Chambers, 2001; Kokubo \& Ida 2007). Similarly, no process other than giant impacts has been shown to be able to successfully tilt the obliquities of Uranus and Neptune (Lee et al., 2007; Morbidelli et al., 2012). Thus, one possibility is that a system of planetary embryos formed by pebble accretion, and that these embryos then collided with each other to form the cores of Uranus and Neptune. 

The number of planetary embryos that form by pebble accretion depends on the number of sufficiently massive seed planetesimals originally in the disk. Kretke \& Levison (2014) performed global simulations of pebble accretion assuming a system of ${\rm\sim100}$ seed planetesimals. In their simulations,  ${\rm\sim100}$  Mars- to Earth-mass planetary embryos form, in a process similar to oligarchic growth. However, the authors observed that these embryos do not merge with each other to form just a few large planetary cores. Instead, they scatter off one another and create a dispersed system of many planets, most of which have a small mass compared to the cores of the giant planets. Moreover, in many of their simulations the system becomes dynamically unstable. Some of the embryos end up in the inner solar system or in the Kuiper belt, which is inconsistent with the current structure of the solar system in these regions. 

In a previous publication (Izidoro et al,. 2015) we showed that the dynamical evolution of a system of planetary embryos changes if the innermost embryo grows into a gas giant planet.  As it transitions from the type I to the type II regime, the giant planet's migration slows drastically such that more distant embryos, also migrating inward, catch up with the giant planet. The gas giant acts as an efficient dynamical barrier to the other embryos' inward migration.  The giant planet prevents them from penetrating into the inner system.  Instead, the embryos pile up exterior to the gas giant. 

We envision the following scenario.  It takes place in a gaseous protoplanetary disk with considerable mass in pebbles.  There is also a population of seed planetesimals.  The two innermost seed planetesimals quickly grew into giant planet cores, achieved a critical mass (Lambrechts et al., 2014) and accreted massive gaseous atmospheres to become Jupiter and Saturn. Jupiter and Saturn do not migrate inward but rather migrate outward (Masset and Snellgrove, 2001; Morbidelli and Crida, 2007; Pierens and Nelson, 2008; Pierens \& Raymond 2011; Pierens et al 2014).  Farther out a number of planetesimals grew more slowly in an oligarchic fashion and generated a system of planetary embryos with comparable masses. The embryos migrated inward until they reached the dynamical barrier posed by the gas giants.  Their mutual accretion produced Uranus and Neptune through a series of mutual giant impacts (and possibly an additional ice giant; see Nesvorny \& Morbidelli 2012), which issued random obliquities for the final planets.  

The goal of this paper is to simulate the late phases of this scenario.  We want to test whether the dynamical barrier offered by Jupiter and Saturn does in fact promote the mutual accretion of these embryos to form a few planet cores.  

A similar study was performed by Jakubík et al., (2012).  In our model we explore a different set of parameters than those considered there. Here, we perform simulations from a wide range of initial numbers and masses of planetary embryos, and adopting different  dissipation timescales for the protoplanetary disk.  Jakubík et al., (2012), instead, restricted the initial planetary embryos to be 3 $M_\oplus$ or smaller. For simplicity they also used a not evolving with time surface density of the gas during their simulations, which covered a timespan of 5 Myr. Thus, our study differs from the previous one by exploring a distinct and more realistic set of parameters.

\subsection{Previous study: Jakubík et al., (2012)}

Before presenting our methods and the results of our simulations we summarize the most important results found  in Jakubík et al., (2012).  We will use them later as a reference for comparison with our results. Using exclusively planetary embryos with initial mass equal to 3 $M_\oplus$ (or smaller) Jakubík et al. (2012) systematically explored the effects of reduced type-I migration rates for the planetary embryos, enhanced surface density of the gas, the presence of a planet trap at the edge of Saturn's gap and of turbulence in the disk.

In the simulations that considered no planet trap, but only a reduced type I migration speed for planetary embryos (with a reduction factor relative to the nominal rate varying between 1 and 6), Jakubík et al. found no significant trends of the results concerning the formation of Uranus and Neptune analogs. They also explored the effects of considering enhanced gas surface densities (scale by a factor up to 6) but despite all considered parameters these simulations still failed systematically in producing good Uranus-Neptune analogs. They usually were able to produce a massive planetary core, larger than ${\rm 10~M_\oplus}$, beyond Saturn; however, the second-largest core on average reached only 6 Earth mass or less. This trend was observed in their entire set of simulations, containing 14 or even 28 planetary embryos of ${\rm 3~M_\oplus}$ each (or smaller). Moreover, the simulations showed that large values for these parameters usually produce massive planets in the inner solar system. However, this high probability of planets crossing the orbit of Jupiter and Saturn and surviving in the inner solar system was most likely overestimated. Such result was presumably a direct consequence of the high gas surface density  assumed for the protoplanetary disks which, in addition, is assumed to remain constant during the 5 Myr integrations (see also Izidoro et al., 2015). 

The Jakubík et al. simulations that considered a planet trap at the edge of Saturn's gap (see also Podlewska-Gaca et al., 2012) marginally increased the mean mass of the largest core. This trend was also observed when enhancing the surface density of the gas. However, in general, the  small mass for the second core remained an issue, as for the simulations without a planet trap. Only one simulation produced two planetary cores of 15 earths masses each beyond Saturn and no other bodies in the inner solar system or on distant orbits.

The Jakubík et al.  simulations considering a turbulent disk (e.g.; Nelson 2000; Ogihara et al 2007) produced typically only one planetary core instead of two. This is because a  turbulent gaseous disk prevents that cores achieve a stable resonant configuration.  Rather, the cores tend to suffer mutual scattering events until they all collide with each other producing a single object. 

All these results were important to help defining the set-up of our simulations. For example, given the weak dependence of the results of Jakubík et al. on many the considered parameters, we assumed in this study the nominal isothermal type I migration rate for the planetary embryos and a gas surface density in the protoplanetary disk equivalent to that of the minimum mass solar nebula (see Morbidelli \& Crida. 2007 and Pierens \& Raymond, 2011).

The structure of this paper is as follow, in Section 2 we detailed our model. In Section 3 we describe our simulations. In Section 4 we present our results, and in Section 5 we highlight our main results and conclusions.

\section{Methods} 

Our study used N-body simulations including the effects of a gaseous protoplanetary disk with a surface density modeled in one dimension (the radial direction; this approach is similar to that of Jakubík et al., 2012 and Izidoro et al., 2015). Although real hydrodynamical simulations would be ideal to study the problem in consideration, there are at two important reasons for our choice. First, hydrodynamical calculations considering multiple and mutual interacting planets embedded in a gaseous disk are extremely computational expensive (eg. Morbidelli et al., 2008; Pierens et al., 2013). It would be impractical to perform  this study using hydrodynamical simulations given the multi-Myr timescale that the simulations need to cover. Second, the method of implementing in a N-body calculation synthetic forces computed from a 1-D disk model is qualitatively reliable. It has been widely tested and used in similar studies, where it was shown to mimic the important gas effects on planets observed in genuine hydrodynamical simulations (e.g. Cresswell \& Nelson, 2006; 2008; Morbidelli et al., 2008).

Our simulations started with fully-formed Jupiter and Saturn orbiting respectively at 3.5 AU and 4.58 AU. This corresponds to the approximate formation location of the gas giants in the Grand Tack model (Walsh et al 2011; Pierens \& Raymond 2011; O'Brien et al 2014; Jacobson \& Morbidelli 2014; Raymond \& Morbidelli 2014).  In practice, the resulting dynamics involved would be only weakly dependent on the orbital radius and the actual range of formation locations for the assumed gas giants is relatively narrow (between roughly 3-6 AU for Jupiter's core). Thus, we do not think that it was worth testing different giant planet's locations.

Beyond the orbit of these giant planets we consider a population of planetary embryos embedded in the gas disk. We performed simulations considering different numbers and masses for the planetary embryos. Here we present simulations considering 2, 3, 5, 10 and 20 planetary embryos. To set the mass of these bodies we define the mass in solids beyond the giant planets which, for simplicity, we call the solid disk mass. We tested two different values for this parameter: 30 and ${\rm 60~M_\oplus}$. This mass is equally divided between the 2-20 migrating planetary embryos. For example, setting the number of migrating planetary embryos equal to 10 and assuming ${\rm 30~M_\oplus}$ in solids, the simulation starts with 10 planetary embryos of ${\rm 3~M_\oplus}$ each. These bodies are randomly distributed beyond the orbits of the giant planets, separated from each other by 5 to 10 mutual Hill radii (e.g. Kokubo \& Ida, 2000). Initially, the eccentricities and inclinations of the embryos are set to be randomly chosen between $10^{-3}$ and $10^{-2}$ degrees. Their argument of pericenter and longitude of ascending node are randomly selected between 0 and 360 degrees. The bulk density of the planetary embryos is set $3g/cm^3$.
 
In our simulations we assume the locally isothermal approximation to describe the disk thermodynamics. Thus, the gas  temperature is set to be a simple power law given by ${\rm T \sim r^{-\beta}}$, where r is the heliocentric distance and ${\rm \beta}$ is the temperature profile index (eg. Hayashi et al., 1981).  We are aware that the direction of type-I migration is extremely sensitive to the disk thermodynamics and to the planet mass (eg. Kley \& Nelson, 2012; Baruteau et al., 2014). Combination of different torques acting on the planet from the gas disk may result in inward or outward migration (Paardekooper \& Mellema 2006; Baruteau \& Masset 2008; Paardekooper \& Papaloizou 2008; Kley \& Crida, 2008; Bitsch \& Kley, 2011, Kretke \& Lin, 2012; Bitsch et al., 2013; 2014). Outward migration is possible only in specific regions of a non-isothermal disk (Kley \& Crida, 2008; Kley et al., 2009; Bitsch et al., 2014). As the disk evolves outward migration must eventually cease.  This is because when the disk becomes optically thin it irradiates efficiently and behaves like an isothermal disk (Paardekooper \& Mellema, 2008). When this happens type-I migration planets simply migrate inward at the type-I isothermal rate (Bitsch et al., 2013; 2014; Cossou et al 2014). Because we assume that Jupiter and Saturn are already fully formed, for simplicity we consider that the disk has evolved sufficiently to behave as an isothermal disk.

\subsection{The gaseous protoplanetary disk}

To represent the gas disk we read the 1-D radial density distribution obtained from hydrodynamical simulations into our N-body code. We assumed a minimum mass solar nebula disk as traditionally used in simulations of the formation of our solar system (Masset et al., 2006; Morbidelli \& Crida, 2007; Walsh et al., 2011; Pierens \& Raymond, 2011).  When performing the hydrodynamical simulations, Jupiter and Saturn were kept on non-migrating orbits and allowed to open a gap in the disk until an equilibrium gas distribution was achieved  (eg. Masset and Snellgrove, 2001). We then averaged the resulting radial profile over the azimuthal direction. Our fiducial profile is shown in Figure 1. In this case, Jupiter is assumed to be at 3.5 AU (its preferred initial location in the model of Walsh et al., 2011) but, as we said above, this is not really important for the results.  In Section 4.5 we will perform simulations with different gap profiles in order to discuss the effects of considering different initial surface density profiles.

In all our simulations the gas disk's dissipation due to viscous accretion and photoevaporation was mimicked by an exponential decay of the surface density, as ${\rm Exp(-t/\tau_{gas})}$, where ${\rm t}$ is the time and ${\rm \tau_{gas}}$ is the gas dissipation timescale. Simulations were carried out considering values for ${\rm \tau_{gas}}$ equal to 1 Myr and 3 Myr. At 3 and 9 Myr, respectively, the remaining gas is removed instantaneously.

In all our simulations the gas disk aspect ratio is  given by 
\begin{equation}
{\rm h=H/r=0.033r^{0.25},}
\end{equation}
where r is the heliocentric distance and H is the disk scale height. 

Still in the hydrodynamical simulation that provide the gas-disk profile, the disk viscous stress is modeled using the standard ``alpha'' prescription for the disk viscosity ${\rm \nu = \alpha c_s H}$ (Shakura \& Sunyaev, 1973), where ${\rm c_s}$ is the isothermal sound speed. In our simulation ${\alpha = 0.002}$.

\begin{figure}[h]
\centering
\includegraphics[scale=.8]{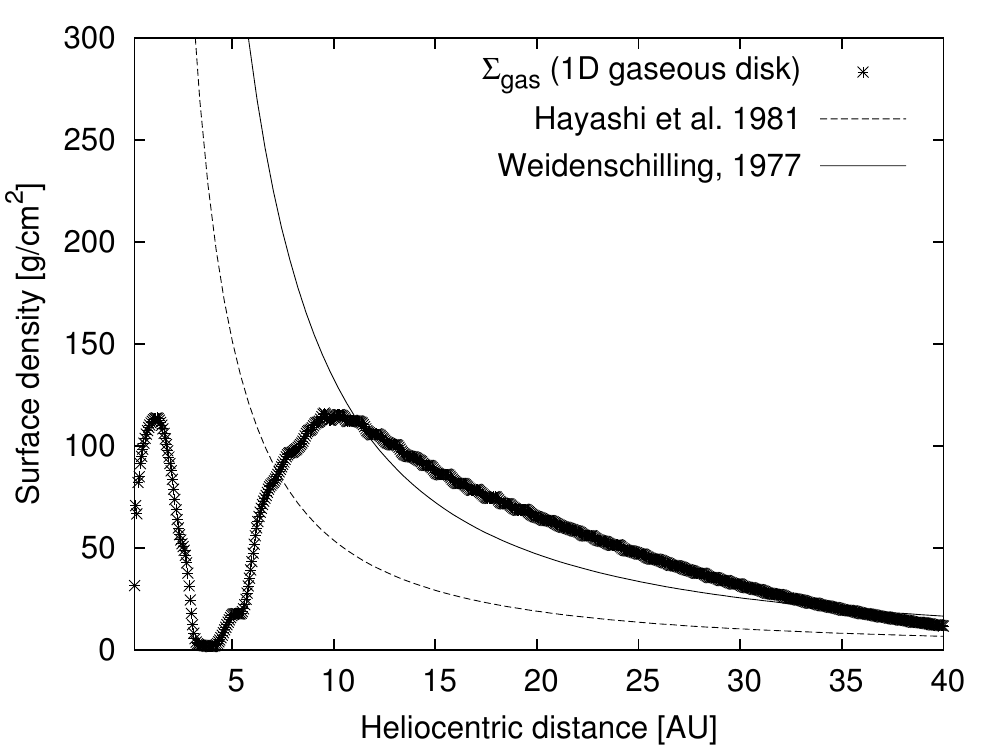}
\caption{Surface density profile generated from a hydrodynamical simulation considering a mininum mass solar nebula and Jupiter and Saturn on fixed orbits. Two variations of the minimum mass solar nebula disk are shown for comparison (Weidenschilling, 1977; Hayashi, 1981).}
\end{figure}

\subsubsection{Tidal interaction of planetary embryos with the gas} 

Our simulations start with planetary embryos of a few ${\rm M_\oplus}$ distributed beyond the orbit of Saturn. These embryos launch spiral waves in the disk and the back reaction of those waves torques the embryos' orbits and makes them migrate (Goldreich \& Tremaine 1980; Ward,1986; Tanaka et al., 2002; Tanaka \& Ward, 2004).  At the same time, apsidal and bending waves damp the embryos' orbital eccentricities and inclinations (Papaloizou \& Larwood 2000; Tanaka \& Ward, 2004).

To include the effects of type-I migration we follow Pardekooper et al., (2011) invoking the locally isothermal approximation to describe the disk thermodynamics. The disk temperature varies as the heliocentric distance as ${\rm T \sim r^{-0.5}}$ and the adiabatic index is set to be ${\rm \gamma=1}$.  In this case,  normalized unsaturated torques can be written purely as function of the negative of the local (at the location of the planet) gas surface density and temperature gradients:
\begin{equation}
{\rm x = - \frac{\partial ln ~\Sigma_{gas}}{\partial ln~r},~~~\beta = - \frac{\partial ln ~T}{\partial ln~r} },
\end{equation}
where ${\rm r}$ is the heliocentric distance and ${\rm \Sigma_{gas}}$  and ${\rm T}$ are the local surface density and disk temperature. Note that the shape of the gas surface density (and consequently the local x) in the region near but beyond Saturn will play a very important role in the migration timescale of planetary embryos entering in this region. Using our surface density profile, x (as in Eq. 2 ) is  dominantly a negative value inside $\sim$ 10 AU. Beyond 10 AU, however, the gas surface density decreases monotonically and x is always positive. Given our disk temperature profile (or aspect ratio) ${\rm \beta=0.5}$

In the locally isothermal limit, the total torque experienced by a low-mass planets may be represented by:

\begin{equation}
{\rm \Gamma_{tot} = \Gamma_L\Delta_{L} + \Gamma_C\Delta_{C}},
\end{equation}
where ${\rm \Gamma_L}$ is the Lindblad torque and ${\rm \Gamma_C}$ represents the coorbital torque contribution. ${\rm \Delta_{L}}$ and ${\rm \Delta_{C}}$ are rescaling functions that account for the reduction of the Lindblad and coorbital torques due to the planet's eccentricity and orbital inclination (Bitsch \& Kley, 2010, 2011; Fendyke 
\& Nelson, 2014). To implement these reductions factors in our simulations we follow Cresswell \& Nelson (2008), and Coleman \& Nelson (2014). The reduction factor ${\rm \Delta_{L}}$  is given as:

\begin{equation}
{\rm \Delta_L = \left[   P_e + \frac{P_e}{|P_e|} \times \left\lbrace 0.07 \left( \frac{i}{h}\right)  + 0.085\left( \frac{i}{h}\right)^4 -  0.08\left( \frac{e}{h} \right) \left( \frac{i}{h} \right)^2 \right \rbrace \right] ^{-1} } ,
\end{equation}
where,
\begin{equation}
{\rm P_e = \frac{1+\left( \frac{e}{2.25h}\right)^{1.2} +\left( \frac{e}{2.84h}\right)^6}{1-\left( \frac{e}{2.02h}\right)^4}}.
\end{equation}

The reduction factor ${\rm \Delta_{C}}$ may be written as:
\begin{equation}
{\rm \Delta_{C}=exp\left(\frac{e}{e_f} \right)\left\lbrace 1-tanh\left(\frac{i}{h} \right)\right\rbrace   },
\end{equation}
where e is the planet eccentricity, i is the planet orbital inclination, and  ${\rm e_f}$ is defined as
\begin{equation}
{\rm e_f = 0.5h + 0.01}.
\end{equation}

Accounting for the effects of torque saturation due to viscous diffusion,  the coorbital torque may be expressed as the sum of the barotropic part of the horseshoed drag,  the barotropic part of the linear corotation torque and the entropy-related part of the linear corotation torque:

\begin{equation}
{\rm \Gamma_C=\Gamma_{hs,baro}F(p_{\nu})G(p_{\nu}) + (1 - K(p_{\nu}))\Gamma_{c,lin,baro} + (1 - K(p_{\nu}))\Gamma_{c,lin,ent}}.
\end{equation}

The formulae for ${\rm \Gamma_L}$, ${\rm \Gamma_{hs,baro}}$, ${\rm \Gamma_{c,lin,baro}}$, and ${\rm \Gamma_{c,lin,ent}}$ are:

\begin{equation}
{\rm \Gamma_L= (-2.5 -1.5\beta + 0.1x)\Gamma_0},
\end{equation}
\begin{equation}
{\rm \Gamma_{hs,baro}= 1.1\left( \frac{3}{2}-x\right) \Gamma_0},
\end{equation}
\begin{equation}
{\rm \Gamma_{c,lin,baro}= 0.7\left( \frac{3}{2}-x\right) \Gamma_0},
\end{equation}
and,
\begin{equation}
{\rm \Gamma_{c,lin,ent}= 0.8\beta \Gamma_0,}
\end{equation}
where ${\rm \Gamma_0=(q/h)^2\Sigma_{gas} r^4 \Omega_k^2}$ is calculated at the location of the planet. Still, we recall that q is the planet-star mass ratio, h is the disk aspect ratio, ${\rm \Sigma_{gas}}$ is the local surface density and ${\rm \Omega_k}$ is the planet's Keplerian frequency.

The functions F, G and K are given in Pardekooper et al. (2011; see their equations 23, 30 and 31). ${\rm p_{\nu}}$ is the parameter governing saturation at the location of the planet and is given by 
\begin{equation}
{\rm p_{\nu} = \frac{2}{3}\sqrt{\frac{r^2\Omega_k}{2\pi\nu}x_s^3}},
\end{equation}
where ${\rm x_s}$ is the non-dimensional half-width of the horseshoe region:

\begin{equation}
{\rm x_s=\frac{1.1}{\gamma^{1/4}}\sqrt{\frac{q}{h}}=1.1\sqrt{\frac{q}{h}. }}
\end{equation}

We stress that when calculating the torques above, we assume a gravitational smoothing length for the planet's potential equal to ${\rm b=0.4h}$.

Following Papaloizou \& Larwood (2000) we define the migration timescale as 

\begin{equation}
{\rm t_m =- \frac{L}{\Gamma_{tot}}},
\end{equation}
where L is the planet angular momentum and ${\rm \Gamma}$ is the torque felt by the planet gravitationally interacting  with the gas disk as given by Eq. (3). Thus, for constant eccentricity, the timescale for the planet to reach the star is given by 0.5$t_m$.

Thus, as in our previous studies (Izidoro et al., 2014, 2015), the effects of  eccentricity and inclination damping were included in our simulations following the formalism of Tanaka \& Ward (2004), modified  by Papaloizou and Larwood (2000), and Cresswell \& Nelson (2006; 2008) to cover the case of large eccentricities. The timescales for eccentricity and inclination damping are given  by  ${\rm t_e}$ and ${\rm t_i}$, respectively. Their values are:

\begin{equation}
\scriptsize
{\rm t_e = \frac{t_{wave}}{0.780} \left(1-0.14\left(\frac{e}{h/r}\right)^2 + 0.06\left(\frac{e}{h/r}\right)^3    + 0.18\left(\frac{e}{h/r}\right)\left(\frac{i}{h/r}\right)^2\right),}
\end{equation}
and
\begin{equation}
\scriptsize
{\rm t_i = \frac{t_{wave}}{0.544} \left(1-0.3\left(\frac{i}{h/r}\right)^2 + 0.24\left(\frac{i}{h/r}\right)^3    + 0.14\left(\frac{e}{h/r}\right)^2\left(\frac{i}{h/r}\right)\right),}
\end{equation}
where
\begin{equation}
{\rm t_{wave} = \left(\frac{M_{\odot}}{m_p}\right)  \left(\frac{M_{\odot}}{\Sigma_{gas} a^2}\right)\left(\frac{h}{r}\right)^4 \Omega_k^{-1}.}
\end{equation}
and ${\rm M_{\odot}}$, ${\rm a_p}$, ${\rm m_p}$, ${\rm i}$, ${\rm e}$ are the solar mass and the embryo's semi-major axis, mass, orbital inclination, and eccentricity, respectively.

To model the damping of semi-major axis, eccentricities and inclinations over the corresponding timescales defined above, we included in the equations of motion of the planetary embryos the synthetic accelerations defined in Cresswell \& Nelson (2008), namely:
\begin{equation}
{\rm \bold{a}_m = -\frac{\bold{v}}{t_m}}
\end{equation}

\begin{equation}
{\rm \bold{a}_e = -2\frac{(\bold{v.r})\bold{r}}{r^2 t_e}}
\end{equation}

\begin{equation}
{\rm \bold{a}_i = -\frac{v_z}{t_i}\bold{k},}
\end{equation}
where ${\rm \bold{k}}$ is the unit vector in the z-direction.

All our simulations were performed using the type I migration, inclination and eccentricity damping timescales defined above. 

\section{Numerical Simulations}

We performed 2000 simulations using the Symba integrator (Duncan et al., 1998) using a 3-day integration timestep. The code was modified to include type-I migration, eccentricity and inclination damping of the planetary embryos as explained above. Physical collisions were considered to be inelastic, resulting in a merging event that conserves linear momentum. During the simulations planetary embryos that reach heliocentric distances smaller than 0.1 AU are  assumed to collide with the central body. Planetary embryos are removed from the system if ejected beyond 100 AU of the central star. 

Our simulations represent 20 different set-ups. They are obtained combining different solid disk masses, initial number of planetary embryos  and gas dissipation timescales. For each set, we performed 100 simulations with slightly different initial conditions for the planetary embryos. That means, we used different randomly generated values for the initial mutual orbital distance between these objects, chosen between 5 to 10  mutual Hill radii.

We have performed simulations considering the giant planets on non-migrating orbits and simulations considering Jupiter and Saturn migrating outward in a Grand-Tack like scenario. Simulations considering Jupiter and Saturn migrating outwards are presented in Section 4.7. In both scenarios, during the evolution of the giant planets their orbital eccentricities can increase up to significantly high values  because of their interaction with an inward migrating planetary embryo trapped in an exterior resonance. Counterbalancing this effect, in our simulations, the eccentricities of the giant planets are artificially damped. Jupiter's eccentricity (and orbital inclinations) are damped on a timescale $e_j/(de_j/dt)\simeq10^4 years$ ($i_j/(di_j/dt)\simeq10^5 years$). This is consistent with the expected damping force felt by Jupiter-mass planets, as consequence of their gravitational interaction with the gaseous disk, calculated in hydrodynamical simulations (e.g. Crida, Sandor \& Kley, 2007). The eccentricity (orbital inclinations) of Saturn are damped  on a shorter timescale, $e_s/(de_s/dt)\sim10^3 years$ ($i_s/(di_s/dt)\sim10^4 years$). We consider these reasonable values since only a partial gap is open by Saturn in the disk (see Figure 1). Thus, this planet should feel a powerful tidal damping comparable to the type-I one (see Eq. 16  and 17). In simulations where Jupiter and Saturn are on non-migrating orbits the eccentricity and inclination damping on these giant planets combined with the pushing from planetary embryos migrating inwards tend to move the giant planets  artificially inward. Thus, we restore the initial position of the giant planets in a timescale of $\sim$1 Myr.  Our code also rescales the surface density of the gas according to the location of Jupiter and as it migrates (see Section 4.7).

\section{Results}

In this section we present the results of simulations considering Jupiter and Saturn on non-migrating orbits. Here, we recall that as in Izidoro et al. (2015), most of surviving planetary cores/embryos  in our simulations stay beyond the orbit of Saturn. In other words, in general, it is rare for planetary embryos/cores to cross the orbit of Jupiter and Saturn, have their orbits dynamically cooled down by the gas effects and survive in the inner regions. We call these protoplanetary embryos the ``jumpers'' (Izidoro et al., 2015).  This result is very different from Jakubík et al. where most of the simulations showed objects penetrating and surviving  in the inner solar system (see discussion in Section 1). However, as mentioned before, this latter result is obviously inconsistent with our planetary system. Thus, in our analysis we reject those simulations that produced jumper planets. After applying this filtering process in  our simulations, for each set  of simulations (20 in total) consisting of 100 simulations we are still left with at least $\sim$60\% of the simulations. In other words, the rate of production of jumper planets in all our simulations has an upper limit of $\sim$40\% (see also Izidoro et al., 2015). 

\subsection{The dynamical evolution}

Figure 2 shows the results of two simulations, which illustrate the typical dynamical evolution of populations of inward migrating planetary embryos.  In these simulations we consider initially 3 and 20 planetary embryos. These objects migrate towards Saturn and are captured in mean motion resonances with the giant planets. Migrating planetary cores pile up into resonant chains (Thommes et al., 2005; Morbidelli et al., 2008; Liu et al., 2014). In systems with many migrating embryos the resonant configurations are eventually broken due to the mutual gravitational interaction among the embryos. When this happens, the system becomes dynamically unstable. During this period, planetary embryos are scattered by mutual encounters and by the encounters with the giant planets. Some objects are ejected  from the system (or collide with the giant planets), while others undergo mutual collisions and build more massive cores. 

 The upper panel (Figure 2) shows a system with just three planetary embryos of ${\rm 10~M_\oplus}$ each. The lower plot shows a system containing 20 planetary embryos of ${\rm 1.5~M_\oplus}$ each. In the simulation with three embryos the system quickly reaches a resonant stable configuration. However, in the system with initially 20 planetary embryos, a continuous stream of inward-migrating embryos generates a long-lived period of instability that lasts to the end of the  gas disk phase. The dynamical evolution of simulations considering an intermediate number of planetary embryos will be shown in Section 4.4.

\begin{figure}
\centering
\includegraphics[scale=.7]{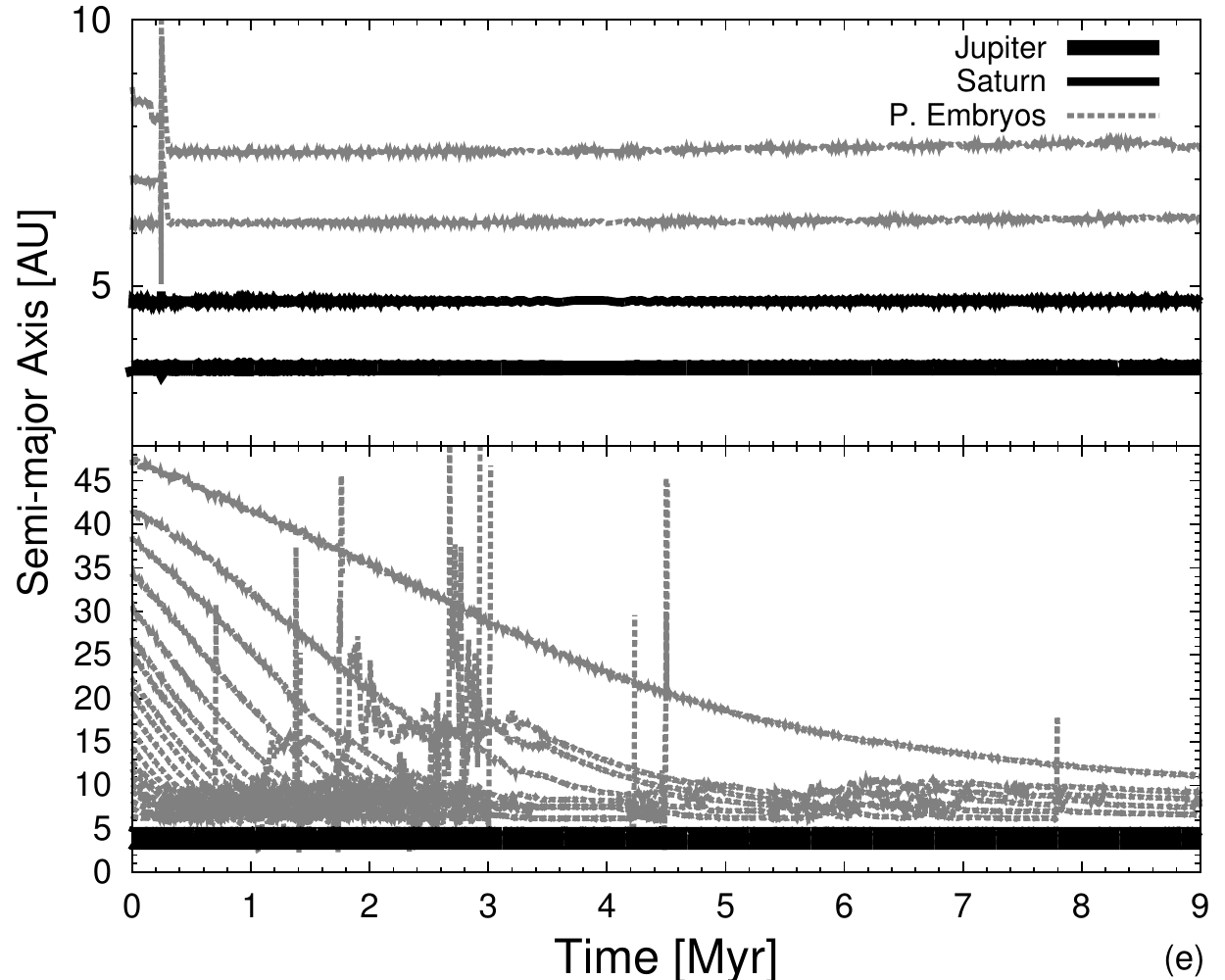}
\caption{Typical dynamical evolution of a population of planetary embryos in two different simulations. In both plots, the horizontal axis represent the time and the vertical one shows the semimajor axis. The upper plot shows the dynamical evolution of three planetary embryos/cores of ${\rm 10~M_\oplus}$ each. The lower plot shows the dynamical evolution of a numerous population of 20 planetary embryos of ${\rm 1.5~M_\oplus}$ each. In both simulations the gas lasts 9 Myr.}
\end{figure}

\subsection{The initial and final number of planetary embryos/cores}

Figure 3 shows the number of surviving embryos/cores as a function of the initial number.  Each dot represents the mean of the results of the 100 simulations in a given series and the vertical error bar represents the maximum and minimal values within the sample over which the mean value was calculated. As expected, there is a clear trend: the more initial embryos, the more survivors. When comparing between sets of simulations with the same initial number of planets but different disk masses or gas disk dissipation timescales, we do not find a clear trend. This is because the initial total mass in protoplanetary embryos considered in  our simulations is only different by a factor of 2 (30 and 60 Earth masses).  In our case, in all set-ups, the simulations that started with 5 planetary embryos ended with a mean of 2-3 survivors. In general, the statistics for the various series of simulations illustrated in Figure 3 are similar. Perhaps the clearest difference is observed for the simulations with 20 initial  embryos. In this case the final number of objects decreases for longer gas dissipations timescales. This is because, when the system starts with as many as 20 objects, 3 Myr (gas lifetime) is not long enough for the system of embryos to reach a final stable configuration with just a few objects (see Figure 2). Actually, even 9 Myr is not long enough and this is why there are still typically more than 5-10 cores in the end. If the number of initial embryos is really that large, the disk's dissipation timescale has to be longer than we considered here.
 
  \begin{figure*}
\centering
\includegraphics[scale=.75]{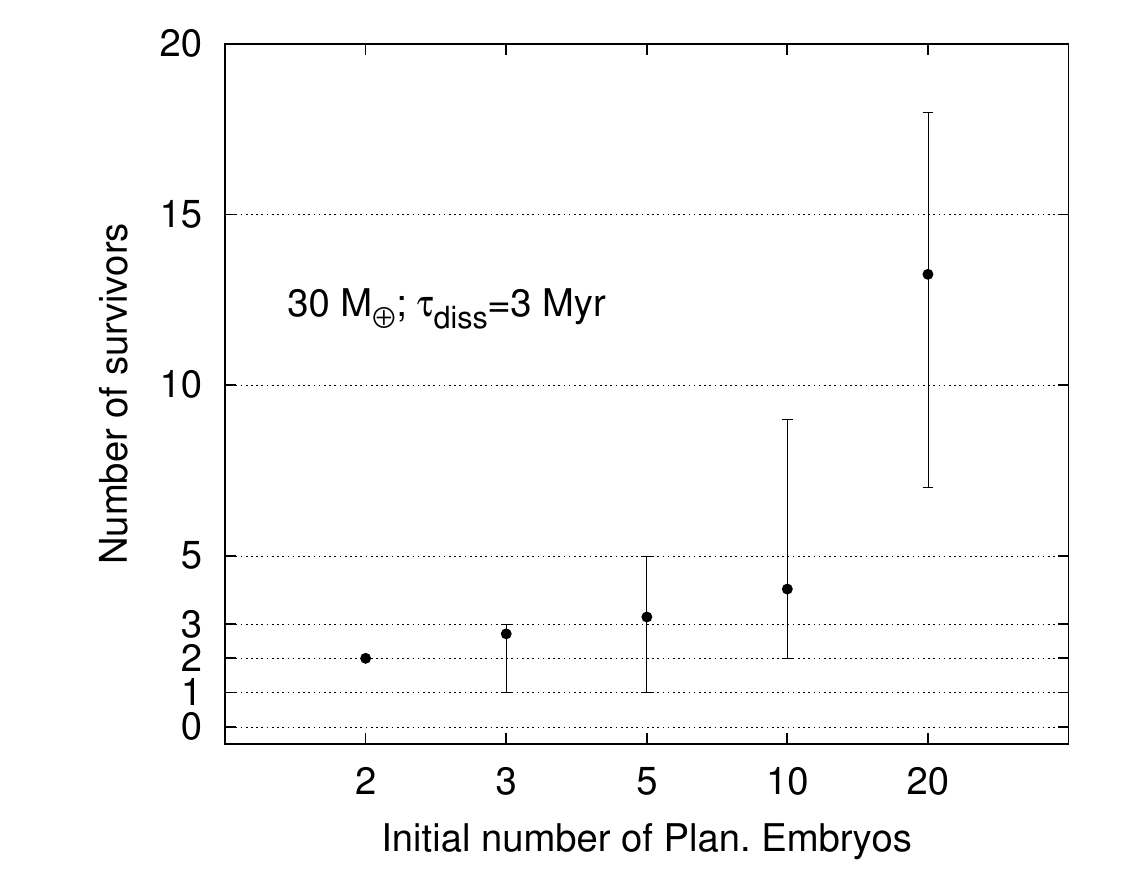}
\includegraphics[scale=.75]{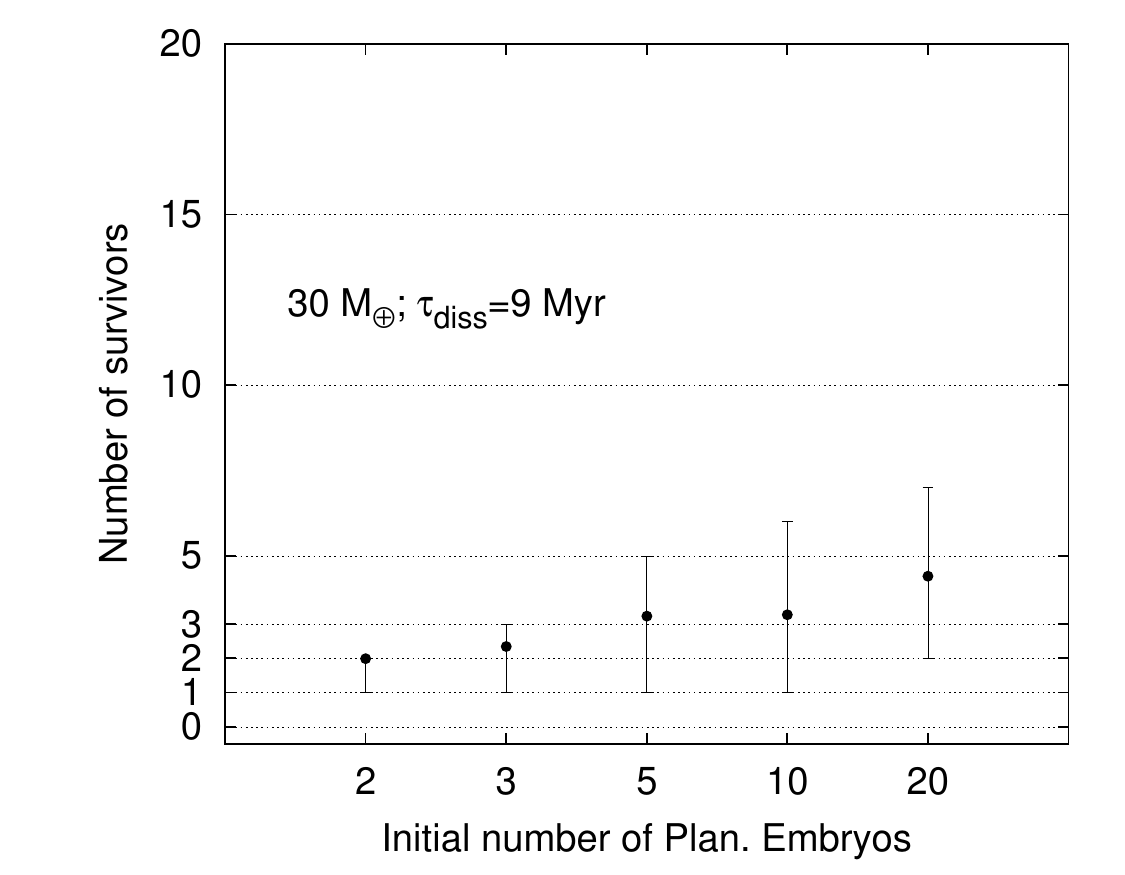}

\includegraphics[scale=.75]{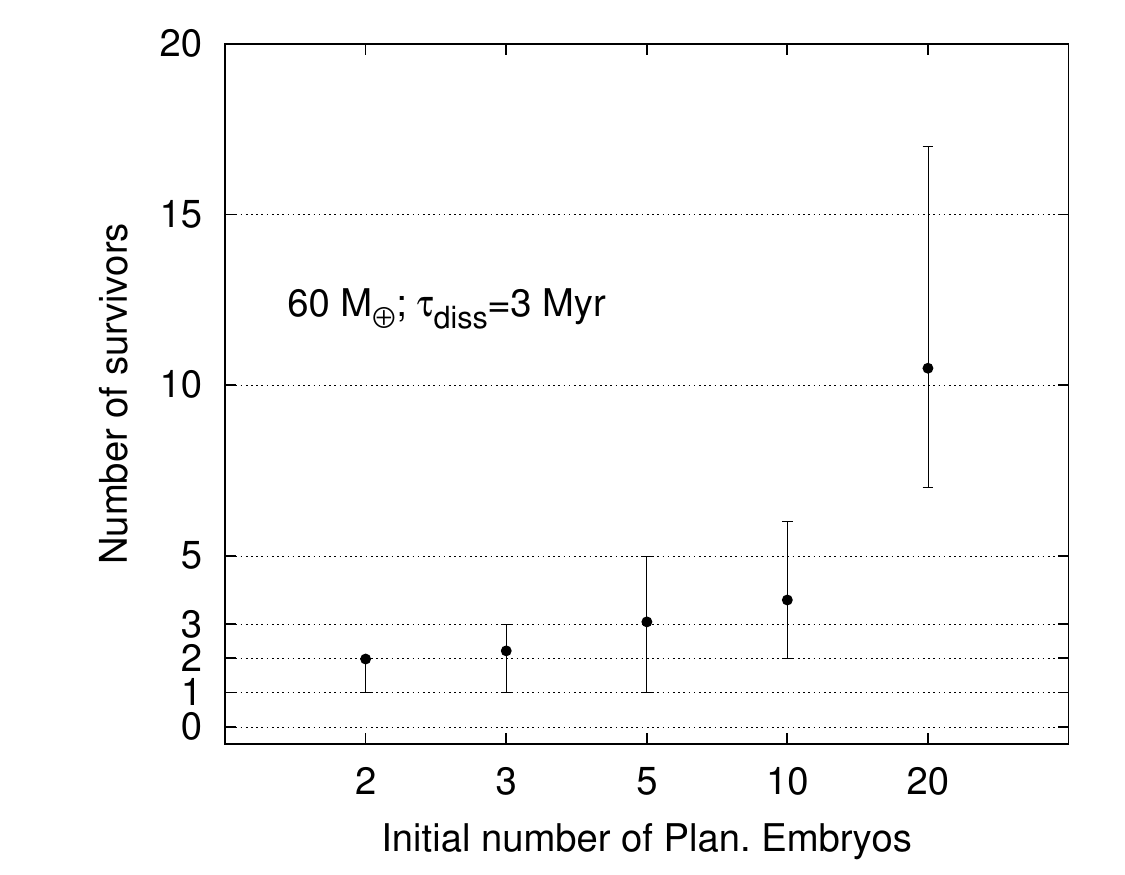}
\includegraphics[scale=.75]{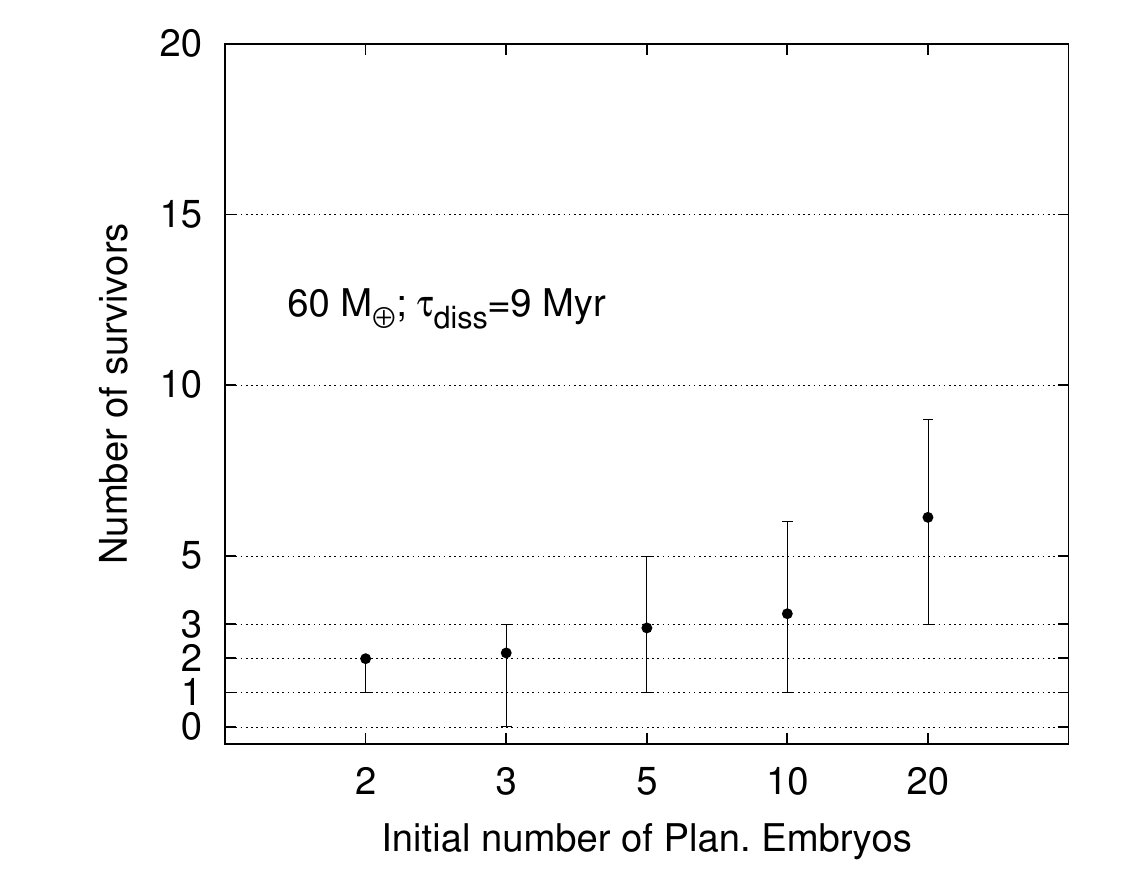}
\caption{Statistical analysis of the results of all simulations. The x-axis shows the initial number of planetary embryos in the simulations. The y-axis shows the final number of cores surviving beyond the orbit of Saturn. The filled circles shows the mean values calculated over those simulations that did not produced jumper planets. The vertical errorbar shows the maximum and minimal values within the sample over which the mean value was calculated. The total initial mass of the disk and the gas dissipation timescale is shown in each panel.}
\end{figure*}

For a given set-up to be consistent with the solar system the final number of cores/embryos should be small: at least two but probably no more than three or four. This is because, numerical models of the dynamical evolution of the outer solar system show that the current architecture could be produced in simulations initially with Jupiter and Saturn plus three or four ice giants, i.e., Uranus, Neptune and a third (possibly fourth) object, all in a compact, resonant configuration (like those we produce here). The rogue planet(s) was eventually ejected from our solar system during the dynamical instability that characterizes the transition from the initial to the current configuration (Nesvorny, 2011; Nesvorny \& Morbidelli, 2012). Thus, just on the basis of the final number of surviving protoplanetary cores (we will consider the issue of mass ratio below), Figure 3 indicates that the best scenarios are those considering between 3 to 10 planetary cores. Consistent with our results, simulations of Jakubík et al., (2012)  considering initially 14 planetary embryos also produced on average between 2 and 3 protoplanetary cores.

The simulations considering initially 2 or 3 planetary objects of 10 $M_\oplus$ or larger demonstrate that it is possible to preserve the initial number of cores if they are not numerous. In this case, however, there would not be giant impacts to explain the large spin tilt that characterize Uranus and Neptune, as discussed in the Introduction.

\subsection{The initial and final masses of the planetary embryos/cores}

Figure 4 shows the final masses of the innermost and second-innermost cores  formed in our simulations (outside the orbit of Saturn). In our set-up, the initial individual masses of the planetary embryos decrease when we increase the number of these objects. Figure 4 shows that, as expected, this property reflects on the final masses of the planets beyond the orbit of Saturn. 
When more than 2 planetary embryos/core survived beyond Saturn, the two largest are in general the innermost ones. When the final number of planetary objects beyond Saturn is larger than 2, the additional ones are, in general, leftover objects that did not grow. Some simulations also produced co-orbital systems (mainly when the gas lasts longer). In these cases, the 1:1 resonant configuration tends to be observed between the innermost (in general, the largest planetary core) and a planetary embryo that did not grow. However, this latter object is not counted as the second innermost in our analysis (see discussion in Section 5).

The masses  of Uranus and Neptune are 14.5 and  ${\rm 17.2~M_\oplus}$, respectively. Figure 4 suggests that it is more likely to produce an innermost planet with about ${\rm 17~M_\oplus}$ in simulations initially with 5 or 10 planetary embryos.  However, the second innermost planet is, in general,  smaller than the innermost one. This also has been observed in simulations by Jakubík et al., (2012). In our simulations, nonetheless, the innermost planetary core is on average $\sim$1.5-2 times more massive than the second one (simulations initially with 5 or 10 planetary embryos) while in Jakubík et al. this number is in general a bit larger, about 2 or 3. The difference with their results is due to three reasons. First because we use a more sophisticated and realistic prescription for the gas tidal damping and migration of protoplanetary embryos. Second, because in our case the gaseous disk dissipates exponentially instead being kept constant over all time. Third, because we have initially different number  and masses for planetary embryos (see how the mass ratio changes depending on these parameters in Figure 4).

 \begin{figure*}
\centering
\includegraphics[scale=.75]{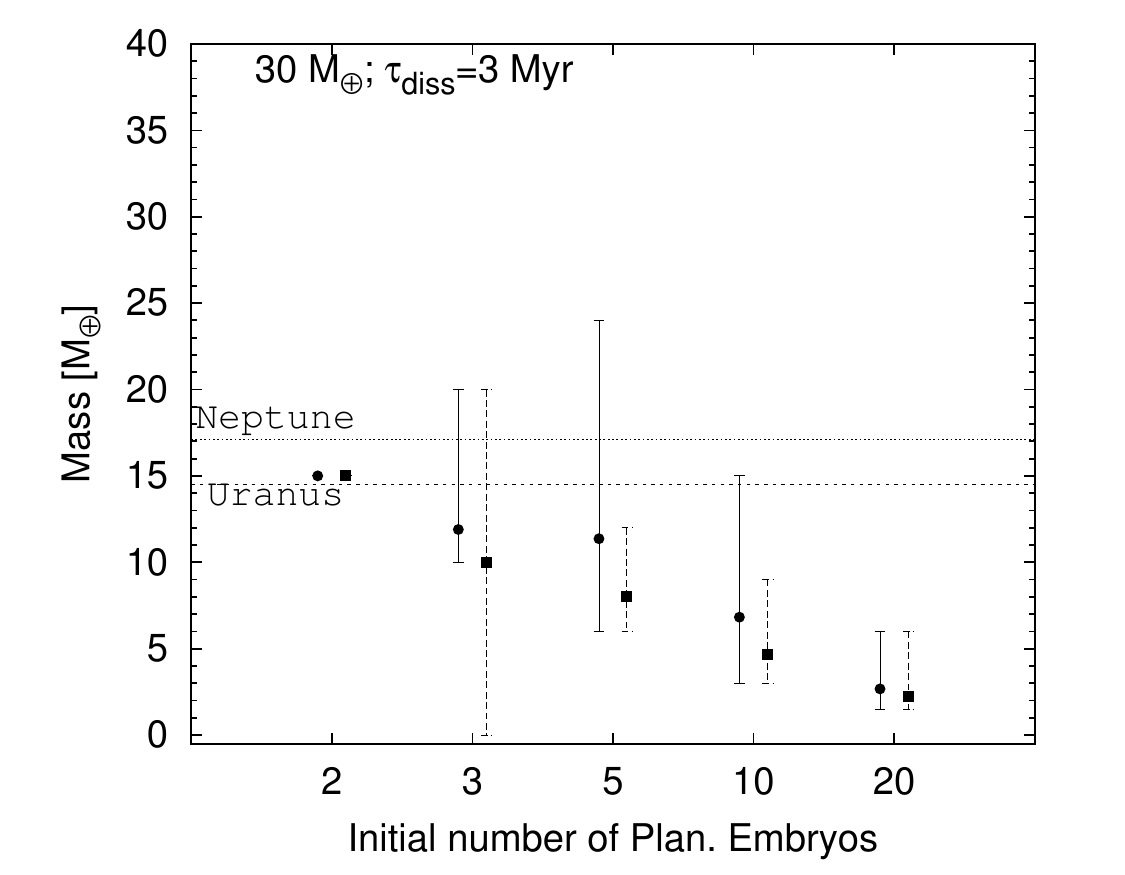}
\includegraphics[scale=.75]{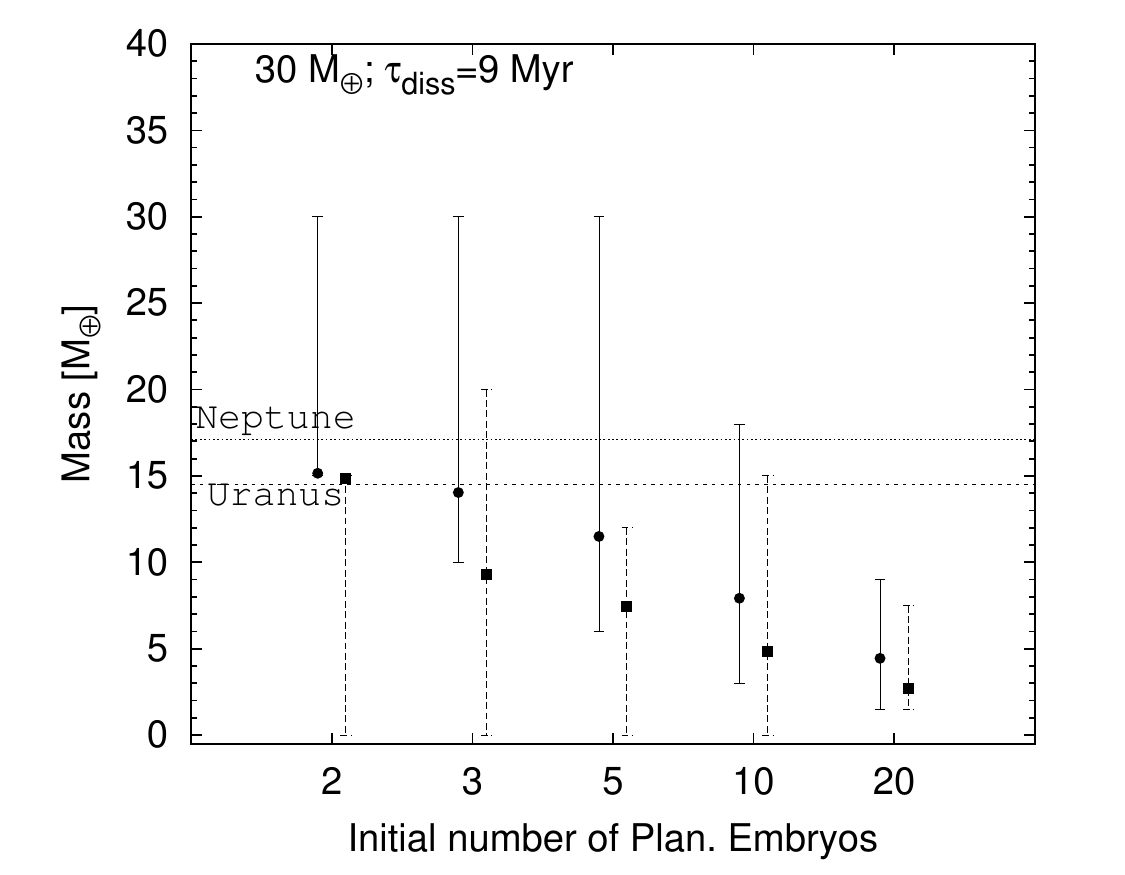}

\includegraphics[scale=.75]{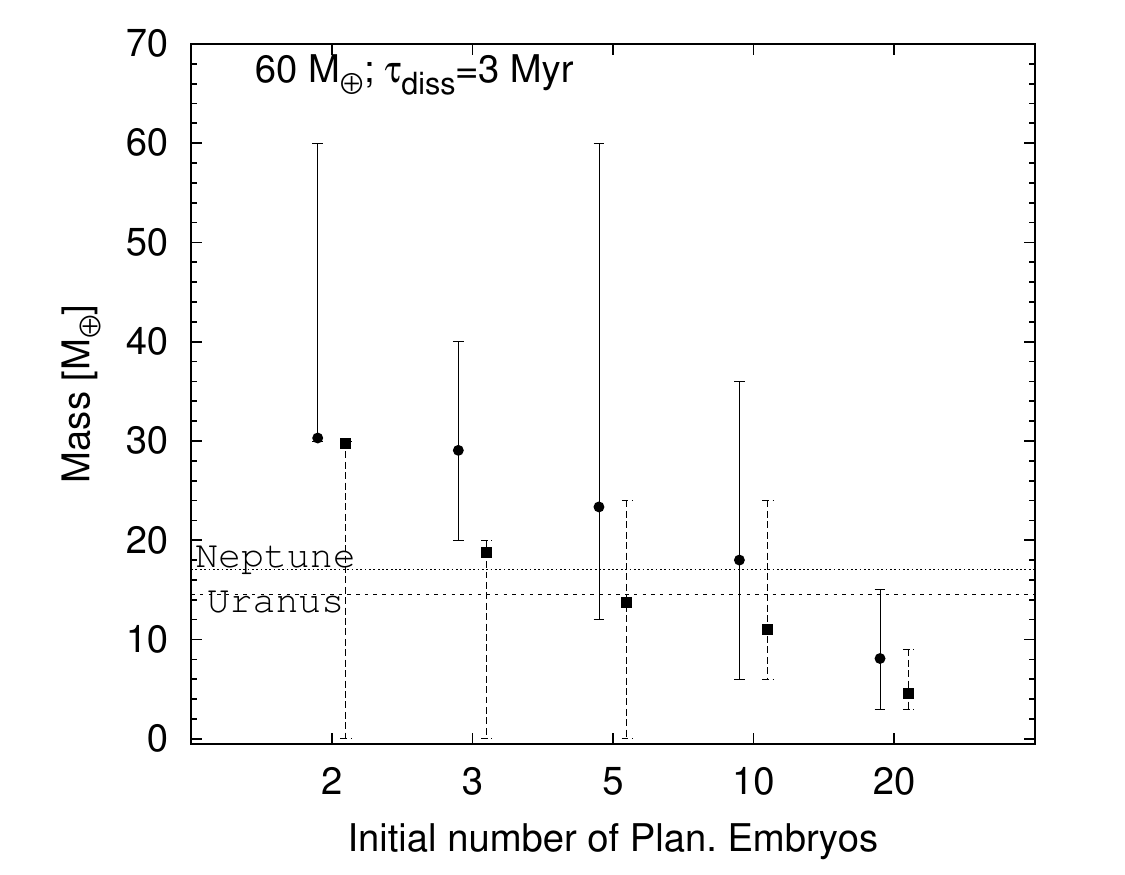}
\includegraphics[scale=.75]{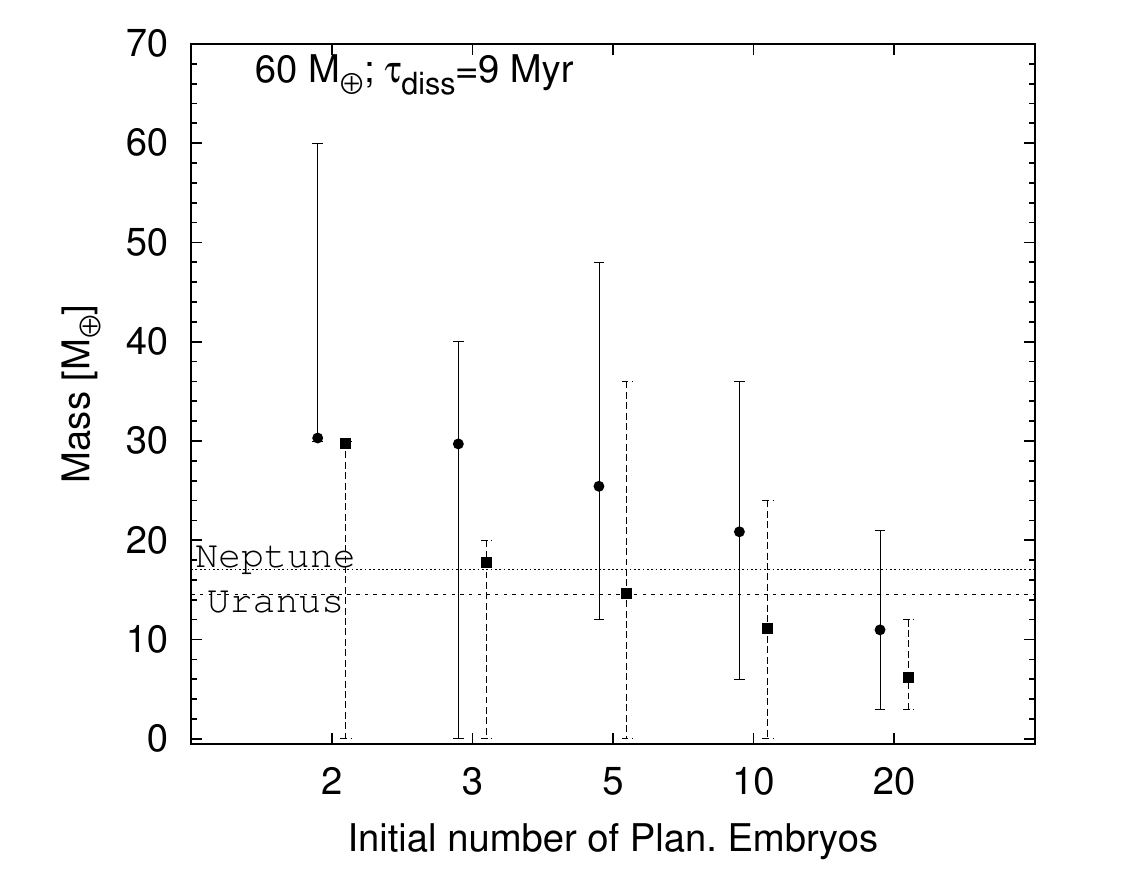}
\caption{Masses of the innermost and second innermost core surviving beyond Saturn for our different sets of simulations.  The filled circles/squares show the mean mass for the innermost and second innermost cores, respectively; the vertical bars range from the maximal to the minimal values obtained.}
\end{figure*}

\subsection{Some of our best results}

We now highlight simulations that formed reasonable Uranus and Neptune ``analogs''. Of course, none of the simulations produced planets with masses identical the ice giants' in our solar system. We do not consider this is a drawback of this scenario but rather a limitation imposed by our simple initial conditions (e.g. all embryos having identical masses). We will present the results of simulations considering  initially planetary embryos  with different masses in Section 4.8~. Also, it is possible that if fragmentation or erosion caused  by embryo-embryo collisions were incorporated in the simulations, it could alleviate this issue and lead to better results. But, we do not expect that these effects would qualitatively  change the main trends observed in our results. 

We have calculated the fraction of the simulations that produced planets similar to Uranus and Neptune. Motivated by the results presented in Section 4.2 and 4.3, we have limited our analysis to those simulations starting with 5 or 10 planetary embryos with individual masses ranging between 3 to  ${\rm 6~M_\oplus}$. This selects eight different sets totaling 800 simulations. We generously tagged a system as a good Uranus-Neptune analog using the following combination of parameters:  (1) both planets beyond the orbit of Saturn (innermost and second-innermost ones) have masses equal to or larger than 12${\rm M_\oplus}$ (i.e. experienced at least one collision each), and (2) their mass ratio\footnote{In the case that the planetary cores have different masses, their mass ratio is defined as the mass of the most massive core (${\rm M_1}$) divided by the mass of the smaller one (${\rm M_2}$).} is ${\rm 1\leq M_1/M_2 \leq 1.5}$ (the mass ratio between Neptune and Uranus is about 1.18).  Please also notice that Neptune is more massive than Uranus, but these planets might have switched position during the dynamical instability phase (Tsiganis et al., 2005).

\begin{table}
\caption{Fraction of sucesss in producing Uranus-Neptune Analogs in two sets of our simulations}              
\label{table:1}      
\centering                                      
\begin{tabular}{|c| c| c| c| c| c|}          
\hline                        
      &  &\multicolumn{4}{|c|}{Gas dissipation timescale}\\ 
      \hline   
T. Mass ${\rm (M_{\oplus})}$ & & \multicolumn{2}{|c|}{3 Myr} & \multicolumn{2}{|c|}{9 Myr}\\   

\hline                                   
       & ${\rm N_{emb}}$ &  5   &  10  &  5     &  10   \\
\hline
    30 & & 25\% & - &  19\%  & - \\      
    60 & & 15\% & 42\% (4\%) & 14\% & 43\% (7\%) \\      

\hline                                             
\end{tabular}
\begin{tablenotes}
      \small
      \item The columns are: initial total mass in protoplanetary embryos (T. Mass  ${\rm (M_{\oplus})}$), number of planetary embryos (${\rm N_{emb}}$), and gas dissipation timescale (3 or 9 Myr). The numbers expressed in percentage report the fraction of simulations which are successful, namely having the mass ratio of the two most massive cores beyond Saturn between ${\rm 1\leq M_1/M_2 \leq 1.5}$ (innermost and second innermost cores beyond Saturn), and each of them have experienced at least one giant collision (their masses are at least as large as ${\rm 12~M_\oplus}$). The values in brackets correspond to the fraction of simulations where at least one of the Uranus-Neptune analogs suffered at least two giant collisions (and the other object, at least one), their mass ratio is between ${\rm 1\leq M_1/M_2 \leq 1.35}$ and they are both at least larger as 12 Earth masses.  
    \end{tablenotes}
\end{table}

 Simulations that satisfied these two conditions produced up to 6 planetary embryos/cores beyond Saturn but in most cases just 3 or 4 objects. In a very small fraction of our simulations that produce Uranus and Neptune analogs ($<10\%$) we did observe the formation of two  planetary cores  where the second innermost one (beyond Saturn) is larger than the innermost one.

About 0\% -- 43\%  of our simulations satisfied the two conditions simultaneously (on mass ratio and individual mass). This shows clearly that the fraction of successful simulations varies depending on the initial number of planetary embryos in the system, their individual and total masses. This may also explain, at least partially, why Jakubík et al., (2012) produced Uranus and Neptune analogs in only one simulation. As discussed before, we explored in this work a much broader set of parameters of this problem.

Figure 5 shows the dynamical evolution of some of the most successful cases. Notice that most of the collisions tend to happen during the first Myr of integration. Moreover,  in general, about 2-3 collisions occur for each planet, which may explain the observed obliquities of Uranus and Neptune (Morbidelli et al., 2012).

In all the simulations illustrated in Figure 5 the accretion of planetary cores is fairly efficient in the sense that the final mass retained in the surviving largest cores is in general about 50\% or so of the initial mass. For example, in the simulation from Figure 5a  the accretion efficiency was of 100\%. In this case,  ${\rm 30~M_{\oplus}}$ in embryos was converted into two cores of 18 and ${\rm 12~M_{\oplus}}$.  All other simulations of Figure 5  show  either the ejection of at least one object from the system,  or collisions with the giant planets, or leftover planetary embryos in the system.

Figure 5-b shows a simulation that formed two planetary cores of ${\rm 24~M_\oplus}$. In this case, each planetary core was formed through two collisions instead three as could be expected given their initial individual masses. First, they hit two planetary embryos growing to 12 Earth masses. Between 0.2 and 0.3 Myr each of these larger bodies hit other two 12 Earth masses bodies reaching their final masses. Note that in this simulation there are two leftover planetary embryos beyond the two largest cores that did not experience any collision.

Figure 5-c shows  one of  our best results compared to the architecture of the solar system. In this case, 10 planetary embryos of ${\rm 6~M_\oplus}$ formed 2 planetary cores of ${\rm 18~M_\oplus}$. Figure 5-d shows one simulation also starting with 10 planetary embryos of 6 Earth masses. In this case  we had also the formation of two planetary cores with  ${\rm 18~M_\oplus}$. However, the third body, in this case, is not a leftover planetary embryo since it has experienced one collision. This is a very atypical result, though. In this case, the gas lasts for 9 Myr and the system  retained the same dynamical architecture for 9 Myr of integration.

Figure 5-e shows another very interesting case where three objects survived beyond the orbit of Saturn. The two innermost objects have masses of 24  and  ${\rm 18~M_\oplus}$, while the third one is a  stranded planetary core that did not suffer any collision. In this case the gas lasted for 3Myr.  Figure 5-e shows a simulation containing initially 5 planetary embryos of 12 Earth masses each. Even in this case, where the planetary embryos are initially very massive (12 Earth masses), we note that two of them suffered one giant collision each.

Figure 5 also shows that most of our simulations ended with more than 2 planets beyond Saturn, typically $\sim$3 (see also Figure 3). Despite Figure 5 shows only a sample of our results it may be considered, in this sense, representative of our results. Importantly, we stress that despite the number of final bodies beyond Saturn seem to support the 5-planet version of the Nice model proposed by Nesvorny (2011) and Nesvorny \& Morbidelli (2012), our results do not directly support that scenario. In fact, in the 5-planet version of the Nice model, the rogue planets has a mass comparable to those of Uranus and Neptune (but see Figure 5d). Here the mass of this extra planet is, in general, much smaller. There are successful 6-planet version of the Nice model with 2 rogue planets of about half the mass of Uranus and Neptune, but they need to be initially placed in between the orbits of Saturn and those of Uranus and Neptune. In our results, instead, the surviving small-mass embryos are always exterior to the two grown cores. It will be interesting to try in the future new multi-planet Nice-model simulations with initial conditions similar to those we build here.

\begin{figure*}
\centering
\includegraphics[scale=.6]{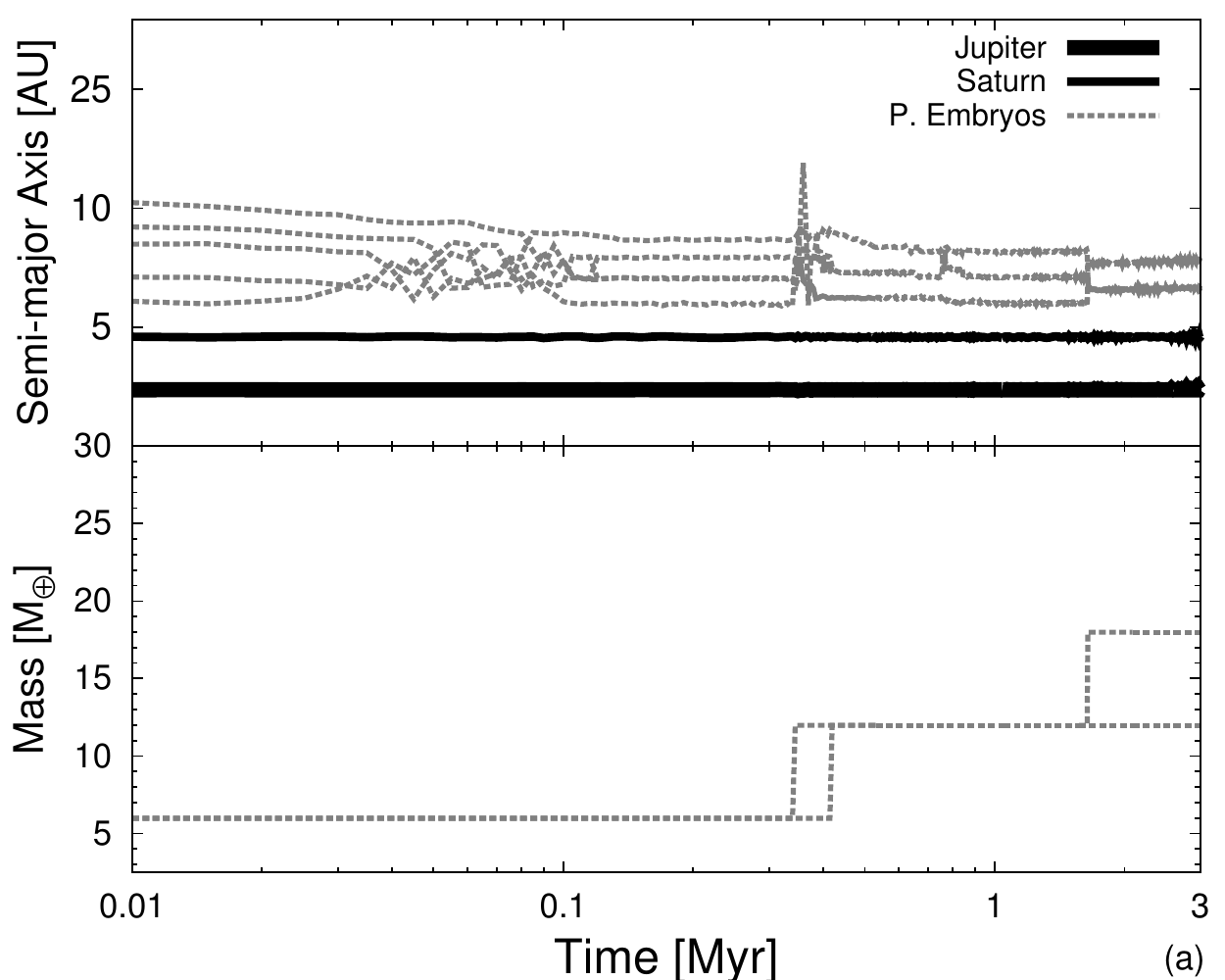}\hspace{0.8cm}
\includegraphics[scale=.6]{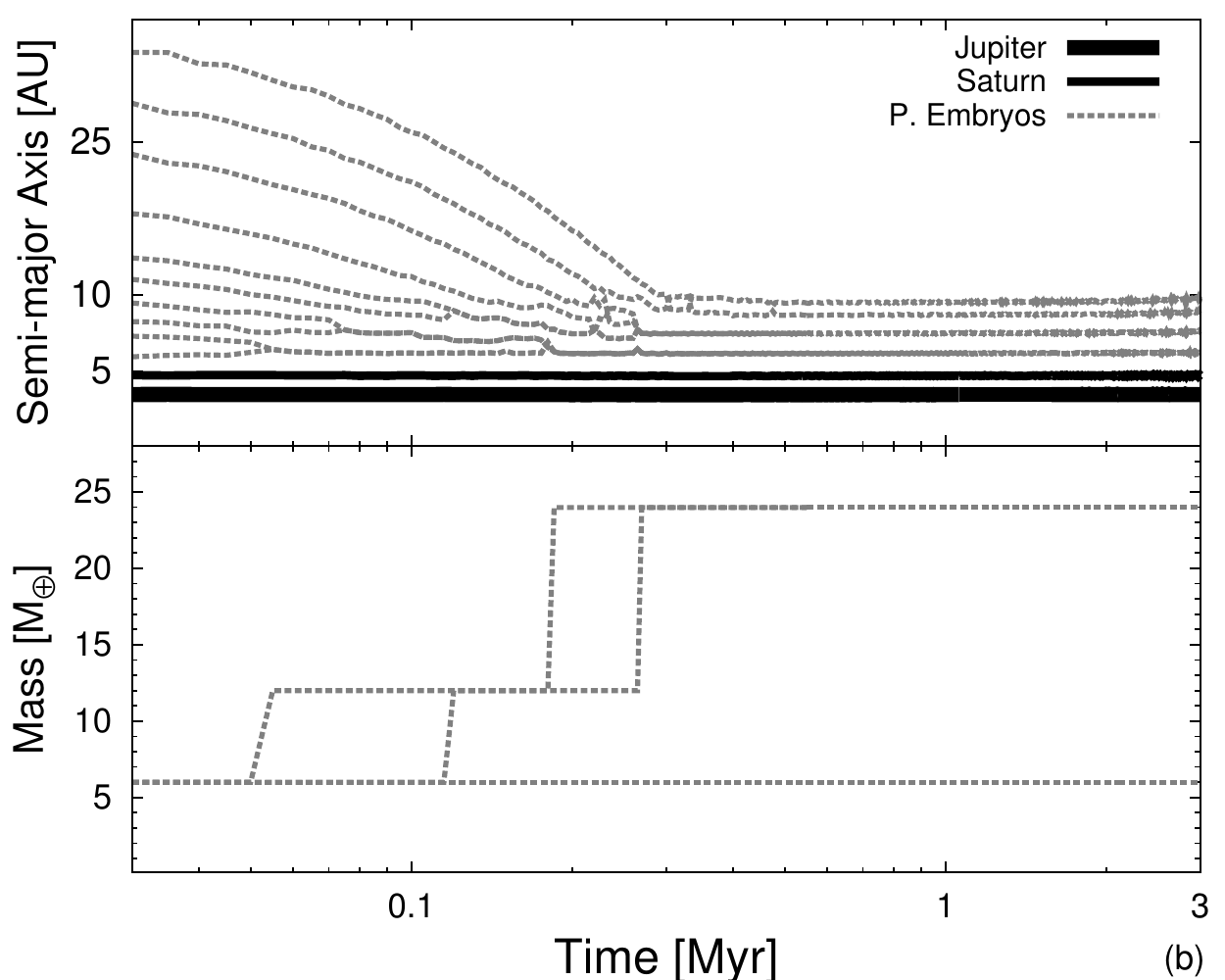}

\includegraphics[scale=.6]{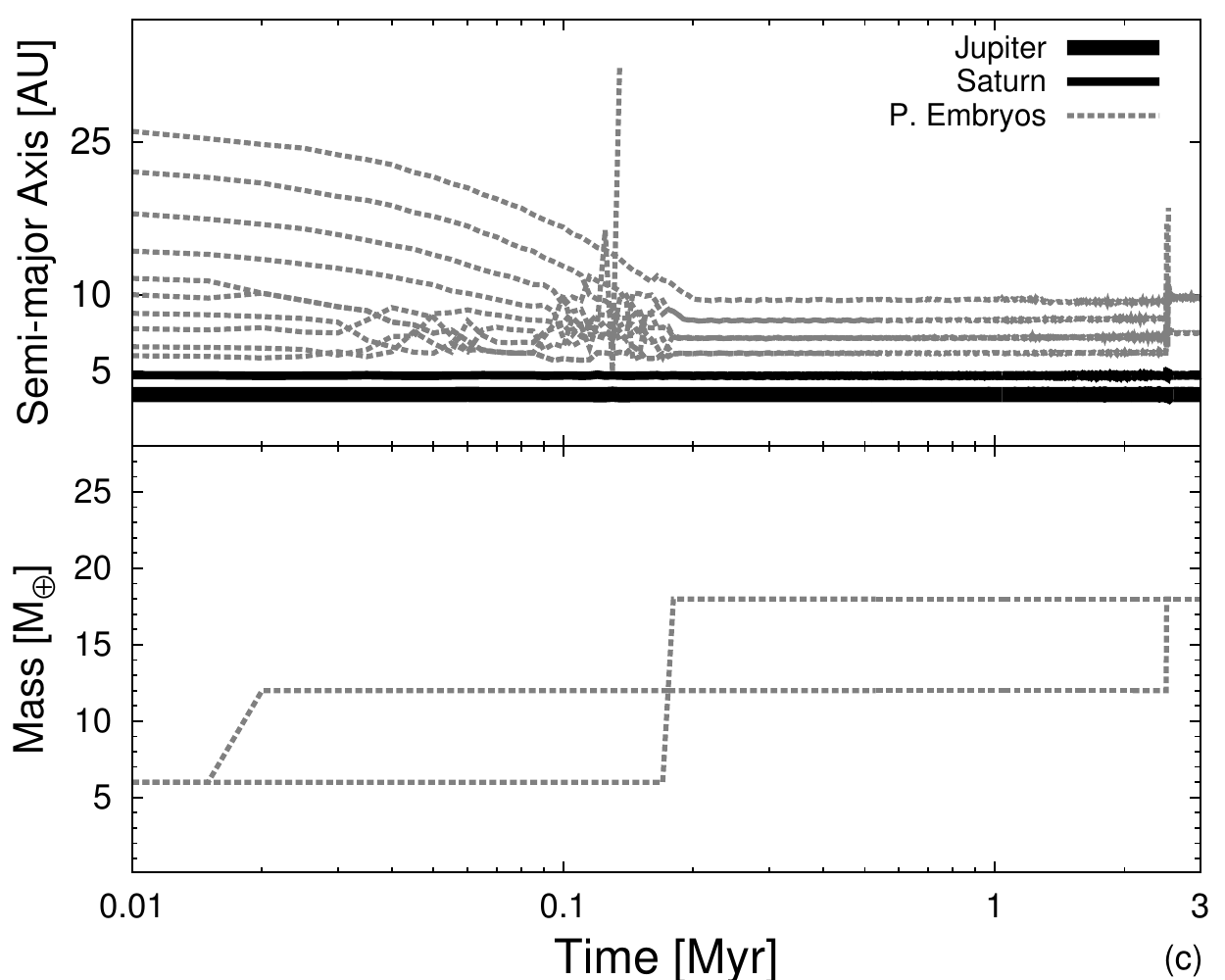}\hspace{0.8cm}
\includegraphics[scale=.6]{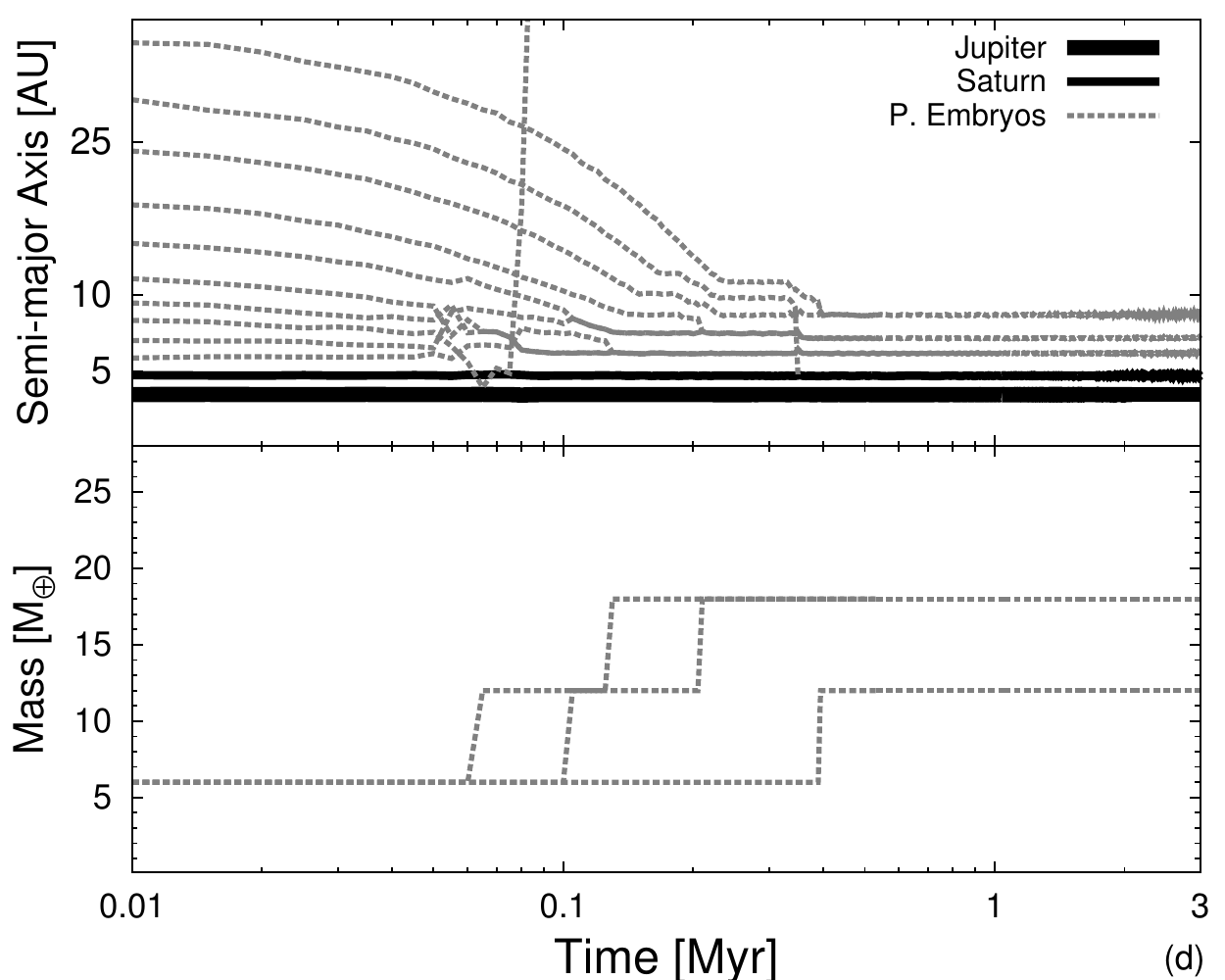}

\includegraphics[scale=.6]{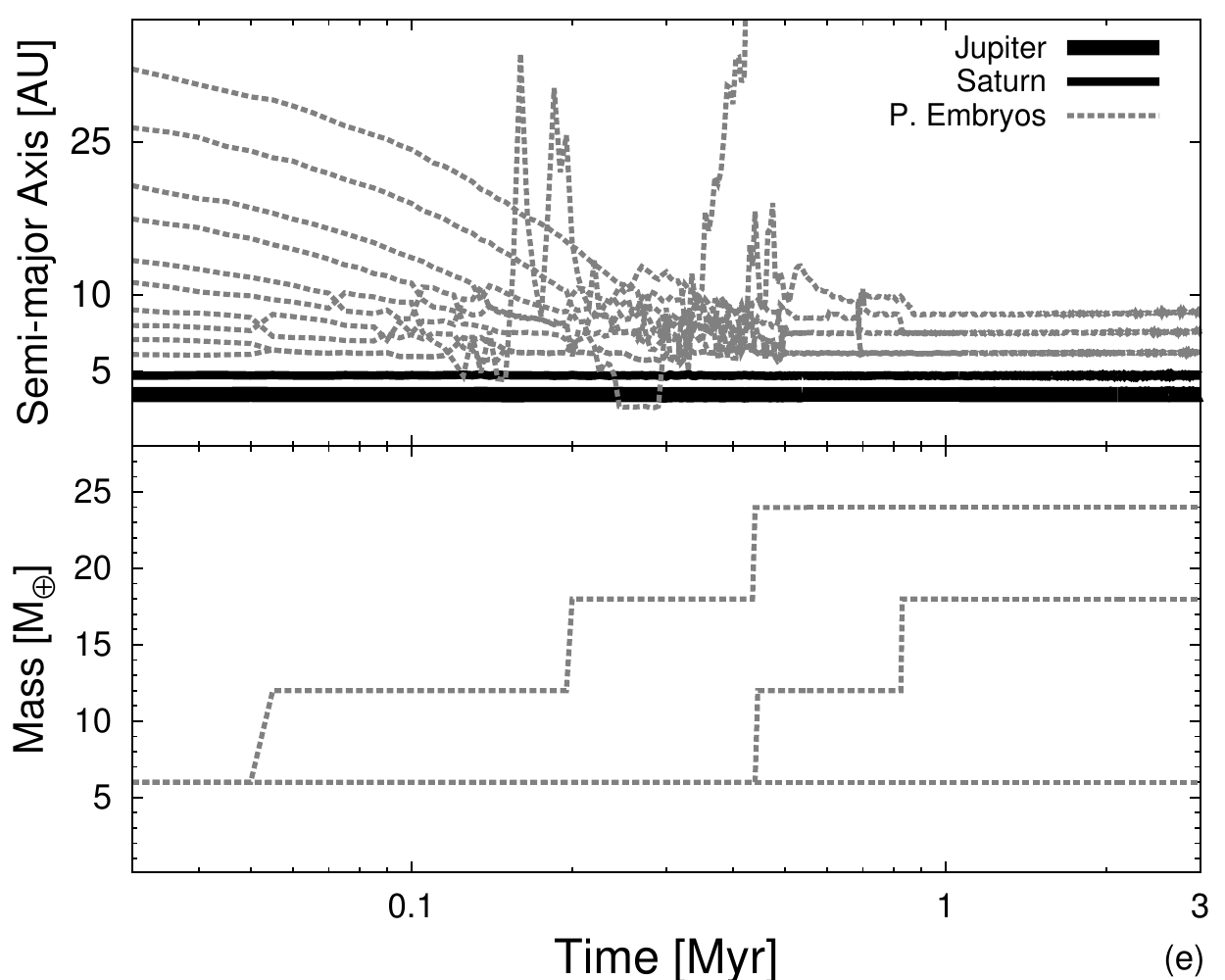}\hspace{0.8cm}
\includegraphics[scale=.6]{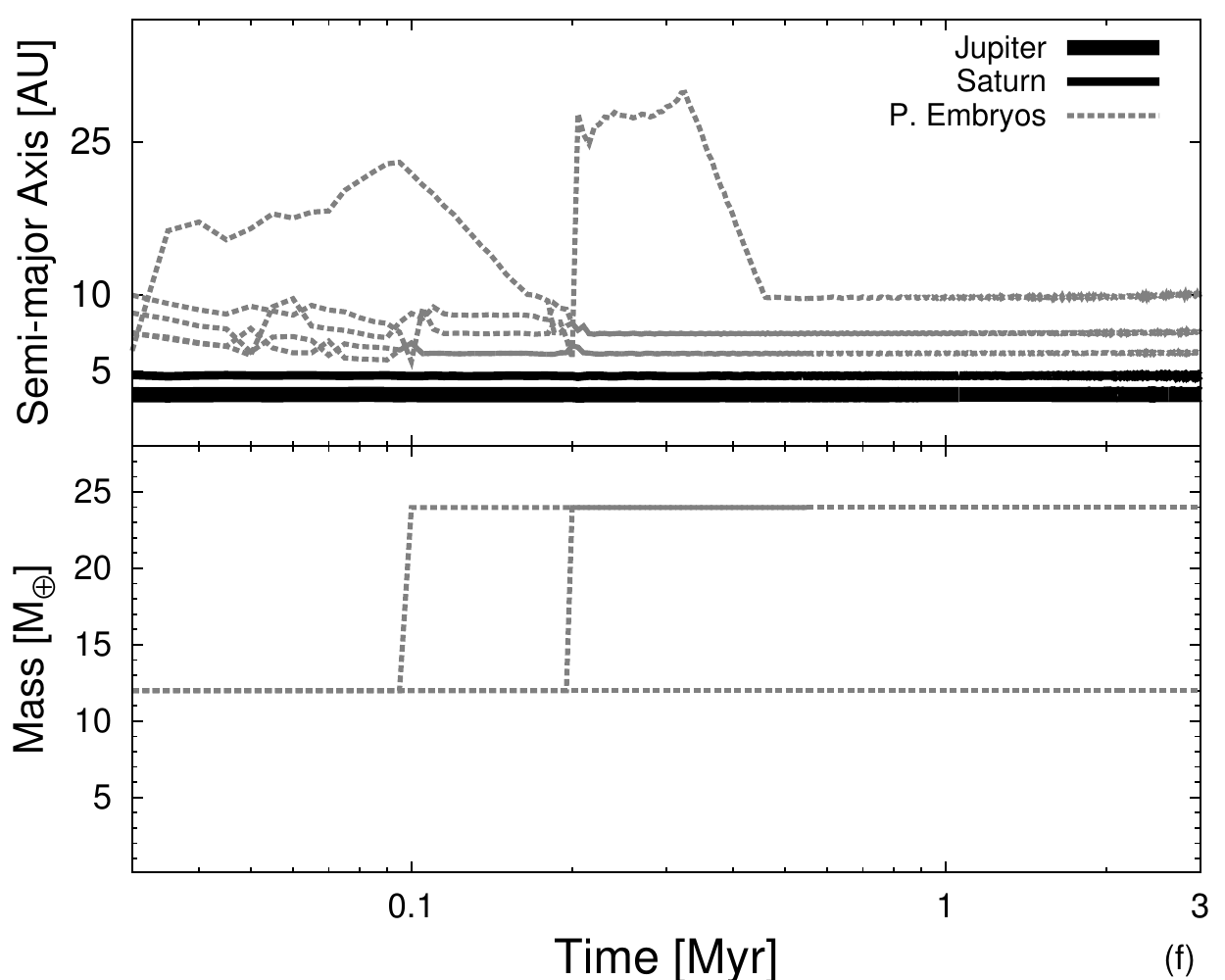}
\caption{Evolution of  planetary embryos leading to the formation of Uranus and Neptune ``analogs'', in six different simulations. Six panels are shown and labelled from a) to f). Each panel refers to a different simulation and is composed by two sticking plots. The upper plot shows the time-evolution of the semi major axis of all migrating planetary embryos (gray) and giant planets (black). The lower plot shows the time-evolution of the mass of those planetary embryos/cores surviving until the end of our integrations.}
\end{figure*}

\subsection{Effect of the initial gas surface density profile}

We also have performed simulations considering different initial gas surface density profiles. For simplicity, we investigate this scenario by rescaling the fiducial gas surface density shown in Figure 1 (${\rm \Sigma_{gas}}$) by a factor ${\rm \epsilon}$. We have assumed values for ${\rm \epsilon}$ equal to 0.4, 0.75,1.2,1.5, and 3. Results of these simulations are summarized in Table 2.

Our results show that a relatively  gas depleted disk is less successful in forming Uranus and Neptune analogs than our simulations with our fiducial gas disk. In a more depleted gaseous disk, planetary embryos migrate slower (towards Saturn) and more often reach and keep, during the gas disk lifetime, mutual stable resonant configurations. Consequently, these simulations tend to have in the end (at the time the gas is gone) more planetary embryos. For example, our simulations considering initially 10 planetary embryos with 6 Earth masses each, and a reduction in the gas surface density given by 60\% ${\rm (0.4\Sigma_{gas})}$ produced on average 5 planets per system (compare with Figure 3). This is also indirectly shown by the mean mass of the innermost and second innermost planetary cores beyond Saturn in Table 2. Note that these objects are systematically smaller when the disk is more depleted. The fraction of success in forming good Uranus-Neptune analogs in these simulations is about 22\%. This shows that the success rate in forming Uranus-Neptune analogs dropped significantly compared to our fiducial model (42\%). In fact,  none of our simulations in this scenario produced two planets beyond Saturn with masses larger than 12 Earth masses, where at least one of them suffered two collisions and their mass ratio is between 1 and 1.35. In this case we also note that the mean mass of the innermost and second innermost planets beyond Saturn are both smaller than 11 Earth masses.

On the other hand, a gas-richer than our fiducial  disk make the planets migrate faster  and this also critically affects the mass ratio between the two innermost planets beyond Saturn. If they migrate inward too fast  it is as if these objects were strongly ``all together'' crunched towards Saturn. This favors that the first innermost planetary core beyond Saturn becomes much more massive than the second innermost one. This for example, tends to reduce the final number of objects in the system. But, consequently, the mass ratio between the first innermost cores beyond Saturn objects tend to increase as is shown also in Table 2. This also leads to a reduction in the success fraction of forming Uranus-Neptune analogs. 

The results presented here show that the  migration timescale of planetary embryos, particularly in the region  very close Saturn where most the collisions happen,  plays a very important role for the  formation of planetary cores with similar masses to those of Uranus and Neptune (or their almost unitary mass ratio).

\begin{table}
\caption{Effects of the initial gas surface density}              
\label{table:1}      
\centering                                      
\begin{tabular}{|c| c| c| c| }          
\hline                        
Scaled surface & Success  & Mean Mass  & Mean mass \\
 density &  Fraction &  innermost   & 2nd innermost \\
\hline
0.4${\rm \Sigma_{gas}}$   &   21\%               & 10.5 (24-6)               & 8.7 (24-6)       \\                   
0.75${\rm \Sigma_{gas}}$  &   33\%                      &  15.7 (30-6)    & 9.5 (18-6)  \\
1.0${\rm \Sigma_{gas}}$ [fiducial]& 42\%  &    18 (36-6)            &  11.3 (24-6)  \\
1.5${\rm \Sigma_{gas}}$   &       37\%          &  21.5 (42-6)  & 11.4 (24-6)\\
3.0${\rm \Sigma_{gas}}$   &       27\%          &   26.7 (48-12) & 12.7 (24-6) \\ 
\hline
\end{tabular}
\begin{tablenotes}
      \small
      \item From left to right the columns are: The scaled surface density, fraction of simulations forming Uranus and Neptune analogs (each core suffered at least one collision, they are both as massive as 12 Earth masses and their mass ratio is between 1 and 1.5), mean masses of the innermost and second innermost planetary cores beyond Saturn. The values in brackets show the range over which the mean values were calculated (compare with Figure 4).
    \end{tablenotes}
\end{table}

\subsection{Obliquity distribution of planets in our simulations}

We tracked the spin angular momentum and obliquity (the angle between rotational and orbital angular momentum of the planet) of the protoplanetary embryos in our simulations assuming that at the beginning our simulations each planetary embryo had no spin angular momentum. When collisions occurred, the spin angular momentum of the target planet was incremented by summing the spin angular momenta of the two bodies  involved in the collision (in the beginning of our simulations they are zero) to the relative orbital angular momentum of the two bodies assuming a two body approximation (eg. Lissauer \& Safronov, 1991; Chambers, 2001). Obviously, this approach assumes that all planetary collisions are purely inelastic and that the star gravitational perturbation may be neglected during the very close-approach between colliding bodies.
\begin{figure}
\centering
\includegraphics[scale=.7]{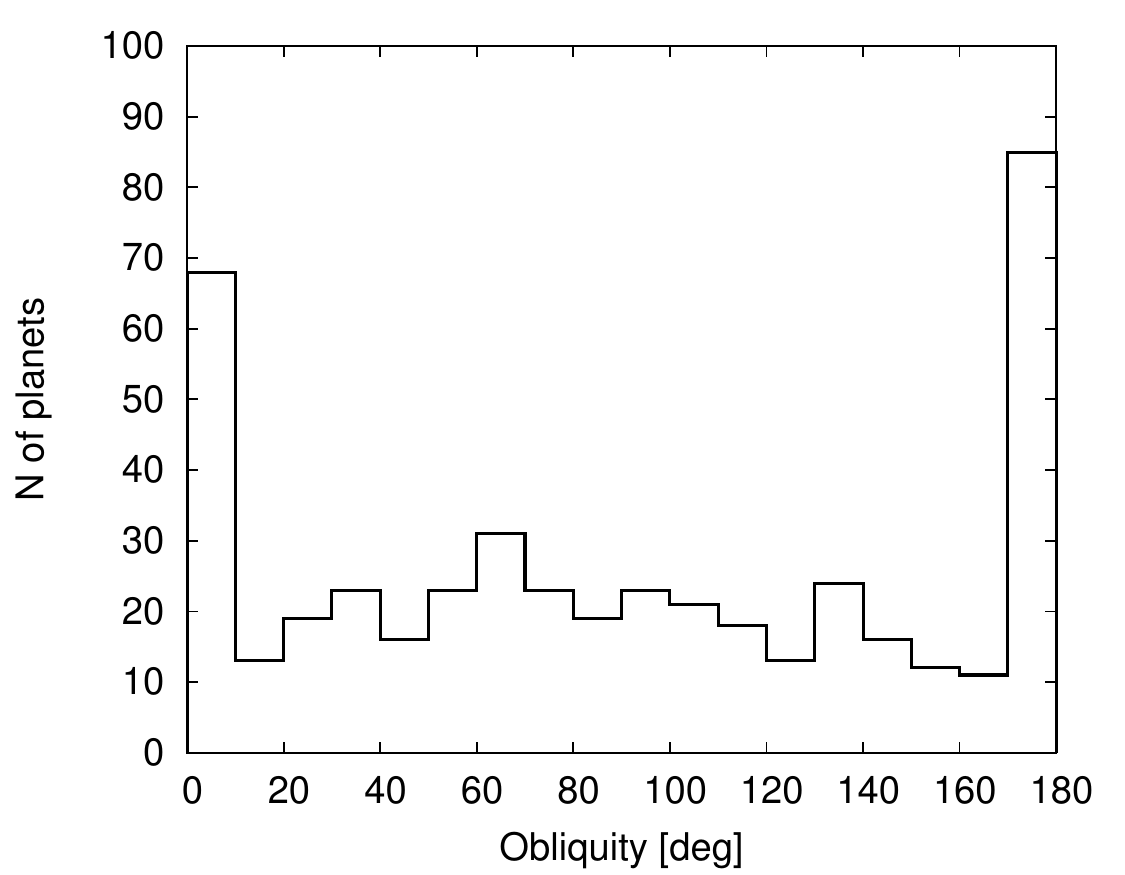}
\caption{Obliquity distribution of all planetary embryos and planets that suffered at least one collision in all our simulations (number initial of planetary embryos = 2, 3, 5, 10 and 20) where the disk total mass is equal to 60${\rm M_{\oplus}}$ and the gas dissipates exponentially in 3 Myr. The vertical axes shows the number of planets and the horizontal one the obliquity.}
\end{figure}

Figure 6 shows the obliquity distribution of the final planets formed in our simulations. The histogram is computed considering only those objects that have suffered at least one collision during the course of our simulations. In this figure, there is clearly a remarkable pile up of bodies either with  obliquities near 0 or 180 degrees (we note that this is a generic trend observed in our results regardless of the initial number of planetary embryos in the system).  However, as also shown in this figure, another significant fraction of this population shows a random distribution between 0 and 180 degrees. This is a very interesting result. The expected distribution of planet obliquities during giant collisions is an isotropic distribution with both prograde and retrograde rotations (Agnor et al., 1999; Chambers, 2001; Kokubo \& Ida  2007). But, different from these previous studies, in our simulations we have the effects of gas tidal damping acting on the planetary embryos which may eventually damp their orbital inclinations to very low values. 

Recall that to tilt (significantly) a planet (target) the projectile needs to hit near the pole of the target. The condition for this to happen is that, at the instant of the physical collision, {\it a}$\times${\it i} > ${\rm {\it R}_{target}}$, where ${ \rm {\it a}}$ is the semi major axis (target and/or projectile), ${\rm {\it i}}$ (radians) is the mutual orbital inclination between projectile and target and ${\rm {\it R}_{target}}$ is the radius of the target\footnote{The relation {\it a}$\times${\it i} > ${\rm {\it R}_{target}}$ assumes, for simplicity, that the projectile and target have circular orbits  and low mutual orbital inclination.}. If we assume for simplicity that: (i) 10 AU is the typical location where our collisions occurs, (ii) that a representative mass of our colliding bodies is about 5 Earth masses, (iii) and these objects have a  bulk density of $\sim$3 g/cm$^3$ and therefore a radius of $\sim$14000 km, we will be in three dimensional collision regime if ${\it i} >{\rm 10^{-5}}$ radians ($\sim$ $6\cdot 10^{-4}$ degrees). In other words, for planets with orbital inclinations below $\sim 6\cdot 10^{-4}$ degrees we should  expected preferentially obliquities near  0 or 180 degrees. Figure 7 shows the obliquity distribution versus orbital inclination and confirm this analysis. The horizontal dashed line in this figure mark the location where {\it i}=$6\cdot 10^{-4}$ degrees. Bodies below this line with obliquities significantly different from 0 or 180 had their orbital inclinations significantly damped by the gas after the giant collisions.

Given our results and the fact that both Uranus and Neptune have large obliquities  suggest that either the tidal damping of the inclinations by the gas-disk was not as strong as in our simulations (eg. the collisions happened when the disk was old and mass starving), or the system was quite crowded of protoplanets (so that there was not enough time to damp the inclinations between mutual encounters), or the disk was turbulent, so that very small inclinations could never be achieved (Nelson, 2005). Among these three alternatives the last one seems to be the most compelling one. The results of our simulations considering a more depleted gas disk (0.4${\rm \Sigma_{gas}}$)  did not show such remarkably pile up of objects with obliquities around 0 and 180 degrees (Figure 6). But, as our results showed, a gas depleted disk tends to decrease the success of forming good Uranus-Neptune analogs.  Moreover, in this scenario, the final systems usually hosts a large number of planetary objects (on average 5). A very numerous population of planetary cores beyond Saturn  would probably make the system dynamically unstable after the gas disk dissipation. Thus, a turbulent disk could be the most elegant solution for this issue (see also Section 5).

\begin{figure}
\centering
\includegraphics[scale=.7]{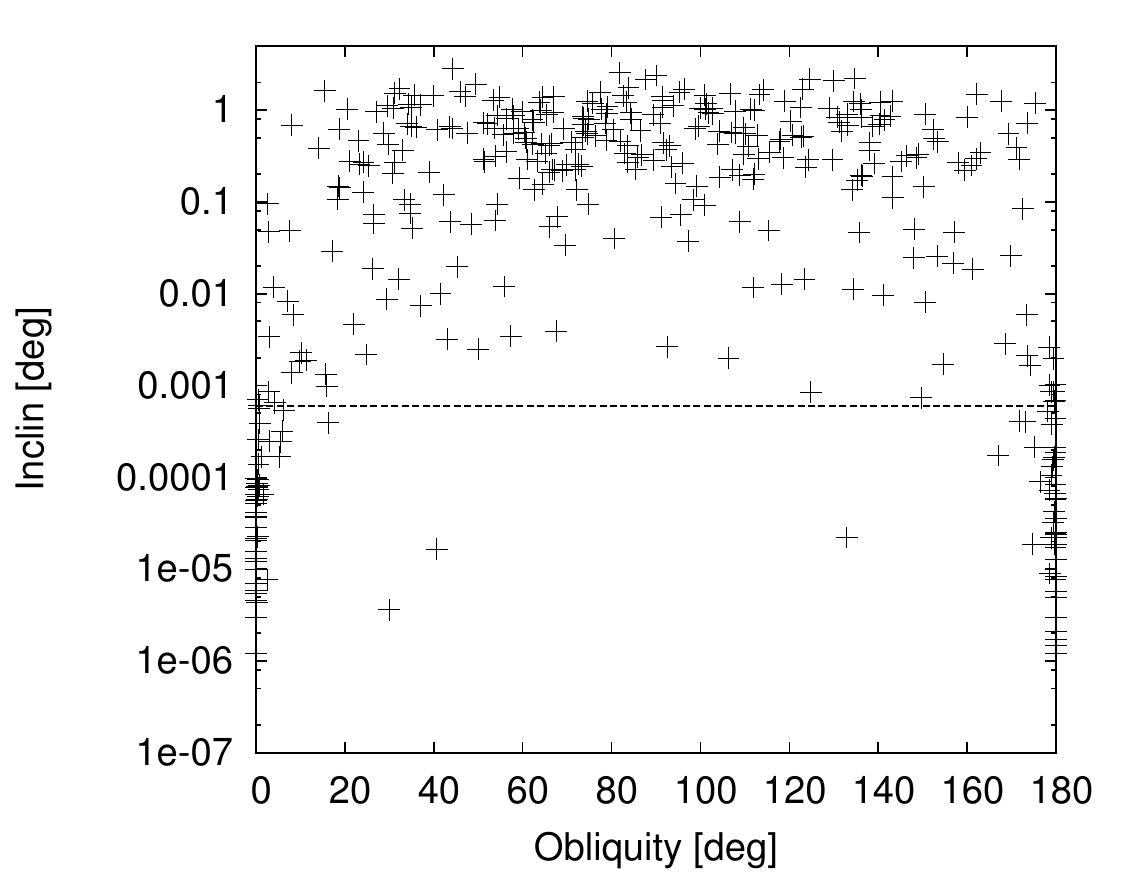}
\caption{Obliquity vs. orbital inclination of all planetary embryos and planets that suffered at least one collision in all our simulations (number initial of planetary embryos = 2, 3, 5, 10 and 20) where the disk total mass is equal to 60${\rm M_{\oplus}}$ and the gas dissipates exponentially in 3 Myr. The dashed line show i=6e-4 degrees}
\end{figure}

\subsection{Effect of Jupiter and Saturn's outward migration}

To this point we have assumed that Jupiter and Saturn are on non-migrating orbits.  The orbital radii of the giant planets were chosen to be consistent with models of the later evolution of the Solar System, specifically the Grand Tack model (Walsh et al., 2011).  But in the Grand Tack model Jupiter and Saturn migrate {\it outward} during the late phases of the disk lifetime.  Outward migration is driven by an imbalance in disk torques which occurs due to the specific Jupiter/Saturn mass ratio and their narrow orbital spacing (Masset \& Snellgrove 2001; Morbidelli \& Crida 2007; Pierens \& Nelson 2008; Pierens \& Raymond 2011; Pierens et al 2014). The question then arises on the effect of the gas giants' outward migration on the accretion of the ice giants.  

We performed additional simulations similar to those presented in section 4 but imposing outward migration of Jupiter and Saturn.  Jupiter and Saturn started at 1.5 and $\sim$2 AU, respectively.  As in Walsh et al (2011) we applied additional accelerations to the planets' orbits to force them to migrate outward.  The gas disk is dissipated exponentially. For the outward migration of Jupiter and Saturn and gas disk dissipation timescales we assumed values consistent with those in Walsh et al. (2011), i.e., ${\rm \tau_{gas}\simeq\tau_{mig}\simeq0.5-1 Myr}$.

\begin{figure}[H!]
\centering
\includegraphics[scale=.6]{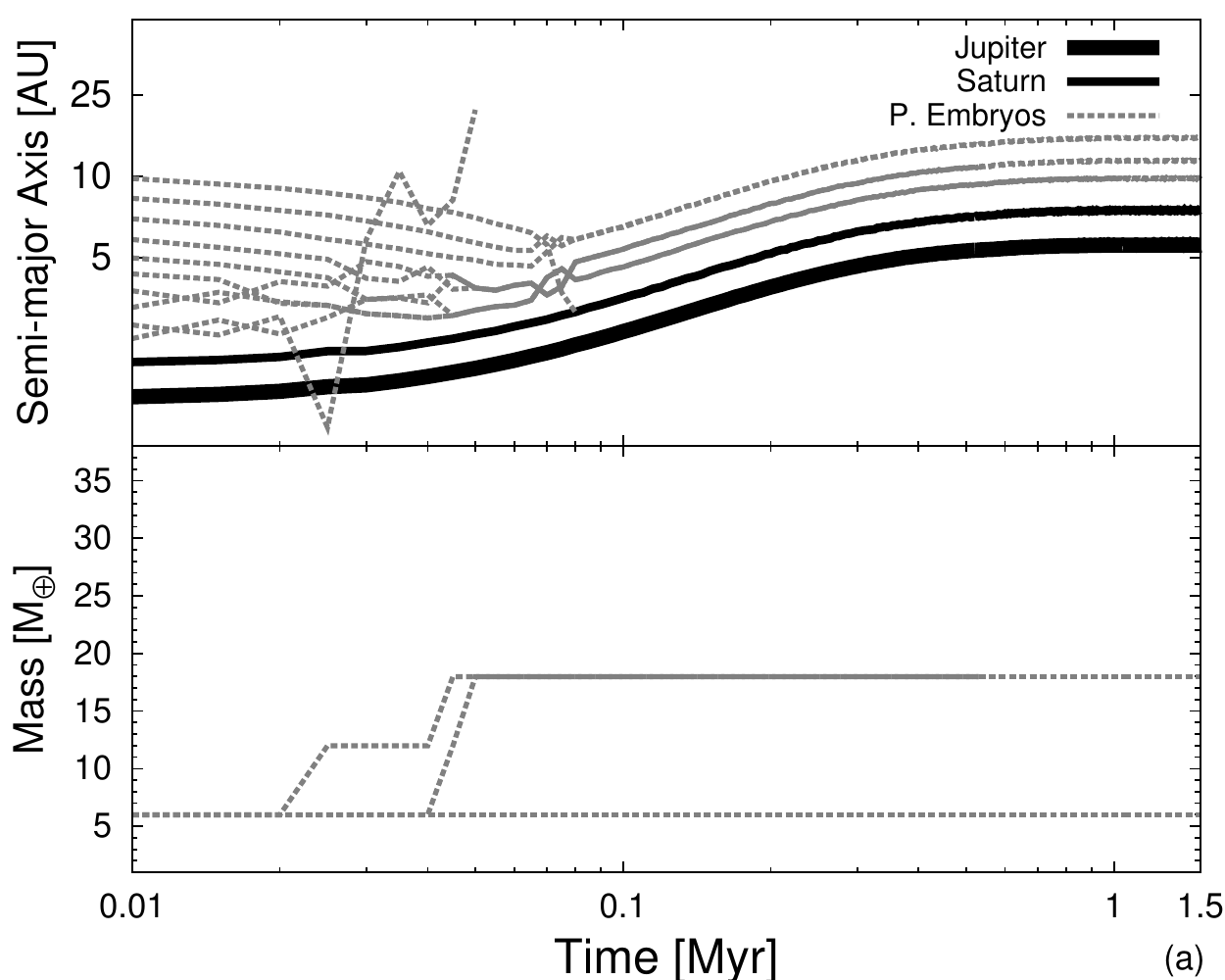}\hspace{0.8cm}
\includegraphics[scale=.6]{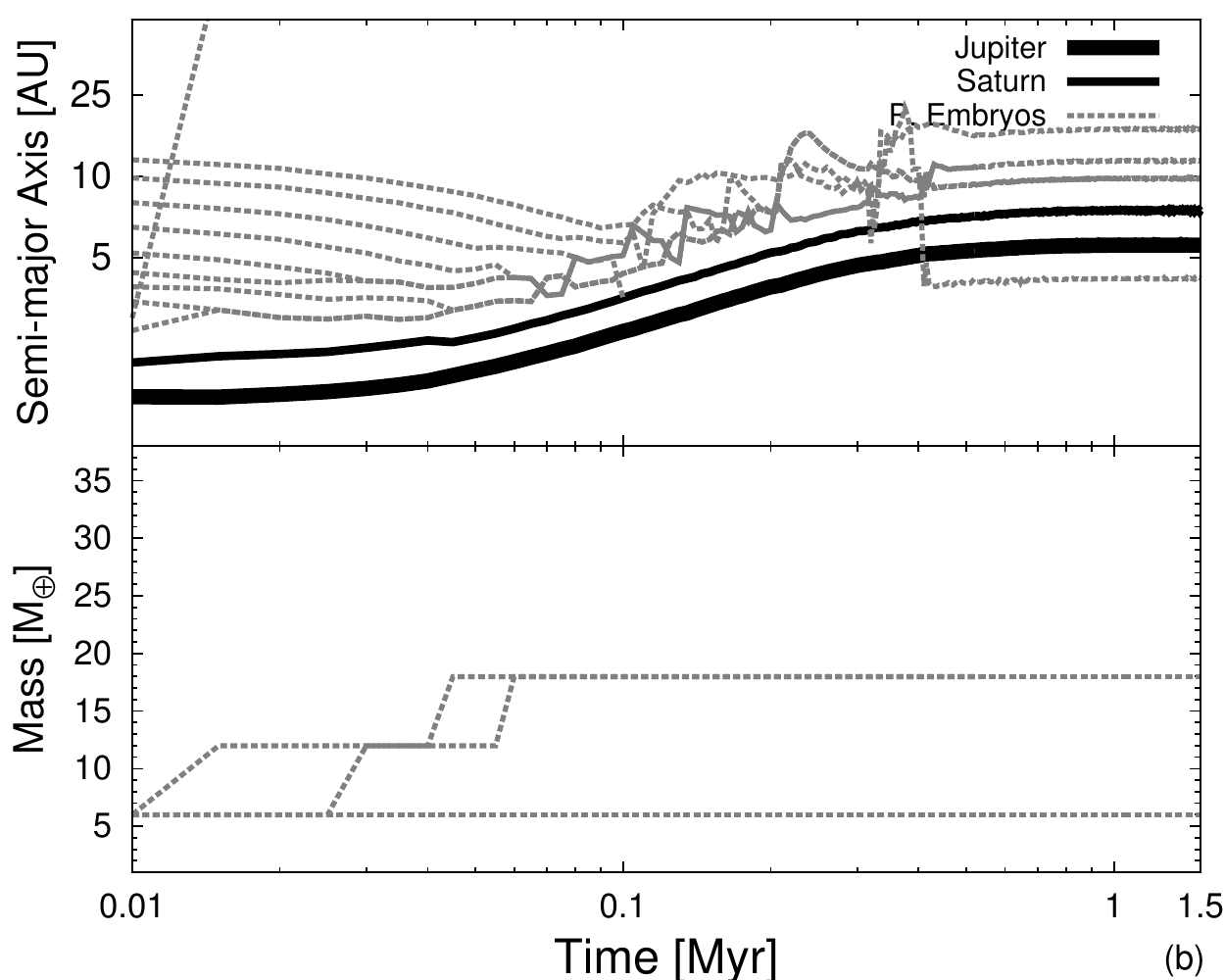}
\caption{Same that Figure 5 but in simulations where Jupiter and Saturn migrate outward. The migration and gas dissipation timescales are 0.5 Myr and 1 Myr, respectivelly. The lower plot show a simulations with a jumper planet. In the lower plot, one of the objects suffered a collision before 0.01 Myr.}
\end{figure}

Figure 8 shows the evolution of one simulation with migrating gas giants.  As in previous simulations, embryos migrate inward, undergo multiple episodes of instability, and pile up in a resonant chain exterior to Saturn. The upper panel in Figure 8 shows a case where there is no jumper planet. Contrasting, the lower panel shows a case where a planet is scattered inward and survives inside the orbit of Jupiter. This certainly makes this simulation inconsistent with the current architecture of our solar system. 

In general,  the main trends observed in those simulations where Jupiter and Saturn are on non-migrating orbits also were observed in simulations considering Jupiter and Saturn migrating outward. Importantly, we stress that  in this scenario it is also relatively challenging to produce two planets with masses comparable to those of Uranus and Neptune. However, simulations with migrating Jupiter and Saturn present some  modest differences relative to those with non-migrating giant planets.

Simulations where Jupiter and Saturn migrate outward tend to produce, on average, less planets than those where Jupiter and Saturn are on non-migrating orbits. For example, in simulations with Jupiter and Saturn migrating outwards and starting with 10 planetary embryos of 6 Earth masses, the final mean number of planetary objects beyond Saturn  is around 2.6 (see Figure 3 for comparison). This is because the outward migration of Jupiter and Saturn combined with the inward migration of the protoplanetary embryos tend to quickly crunch the system into a small region (see Figure 8). During this phase, resonant configurations among these objects and giant planets (or other planetary embryos) tend to be easily broken down. As a result planetary embryos get dynamically unstable, are ejected, scattered inward or suffer mutual accretion. This process repeats until the migration of Jupiter and Saturn is completed. Consequently, the mutual accretion among protoplanetary cores tends to be accelerated and generally happens very early ($\lesssim$ 0.1-0.5 Myr -- e.g. Figure 8).

We also observed that the rate of ``jumpers'' was  higher with migrating giant planets (see Izidoro et al., 2015). For example, for ${\rm \tau_{gas}\simeq\tau_{mig}\simeq0.5-1 Myr}$ and in simulations considering initially 10 planetary embryos with 6 Earth masses each show a rate of jumper of about $\sim$ 50-80\% (depending on the combination between the parameters ${\rm \tau_{gas}}$ and ${\rm \tau_{mig}}$).  This makes sense for two reasons. First, because a higher relative migration rate between the gas giants and embryos should produce stronger instabilities (see Izidoro et al 2014). Second, in simulations where Jupiter and Saturn migrate outward they start  closer to the star (Jupiter starts at $\sim$1.5 AU and Saturn at $\sim$ 2.0 AU). Our code rescales the surface density  of the gas according to the location of Jupiter and as it migrates. Thus, as the closer Jupiter is to the star, the higher is the gas surface density inside its orbit (Walsh et al., 2011) and Izidoro et al., (2015) found that the probability that a planetary embryo jumps across giant planet orbits increases with the gas density. However, the fraction of simulations that produced ice giant analogs with comparable masses is similar in the cases with migrating and non-migrating giant planets.  In fact, the fraction of Uranus and Neptune analogs is  12\% in simulations considering initially 5 planetary embryos of 6 Earth masses each. In simulations considering initially 10 planetary embryos of 6 Earth masses each this number is about 12\% (4\%). Thus the difference in success rates between the simulations with and without migrating giants is not critically different. The values shown in brackets show the fraction of our simulations where at least one of the planetary cores experienced two collisions, both have masses larger or equal to 12${\rm M_{\oplus}}$ and the mass ratio between them is between 1 and 1.35.

\subsection{Simulations with different initial mass for planetary embryos}

 We also have performed 600 simulations considering initially planetary embryos with different masses. To set the mass of these bodies we keep the total mass of disk fixed (30 or 60 Earth masses as in our fiducial model) and we continue inserting  planetary embryos in the system until the set mass limit is reached. We have performed two sets of simulations varying the width of the  distribution of mass of individual planetary embryos. In the first one (hereafter called V1) we have allowed a very wide range of mass,  where the initial mass of the embryos is randomly chosen to be between 1 and 10 Earth masses. In the second one (hereafter called V2), the individual mass of the planetary embryos is randomly chosen in a smaller range, between 3 and 6 Earth masses.

Figure 9 shows two examples of these simulations. Observe that the initial masses of the protoplanetary embryos are different. In both cases, two Uranus and Neptune analogs are formed where the mass of the two largest planetary cores are very similar to those of the real planets. Figure 9-a represents a simulation of the setup V1 while Figure 9-b corresponds to the setup V2.

\begin{figure}[H!]
\centering
\includegraphics[scale=.6]{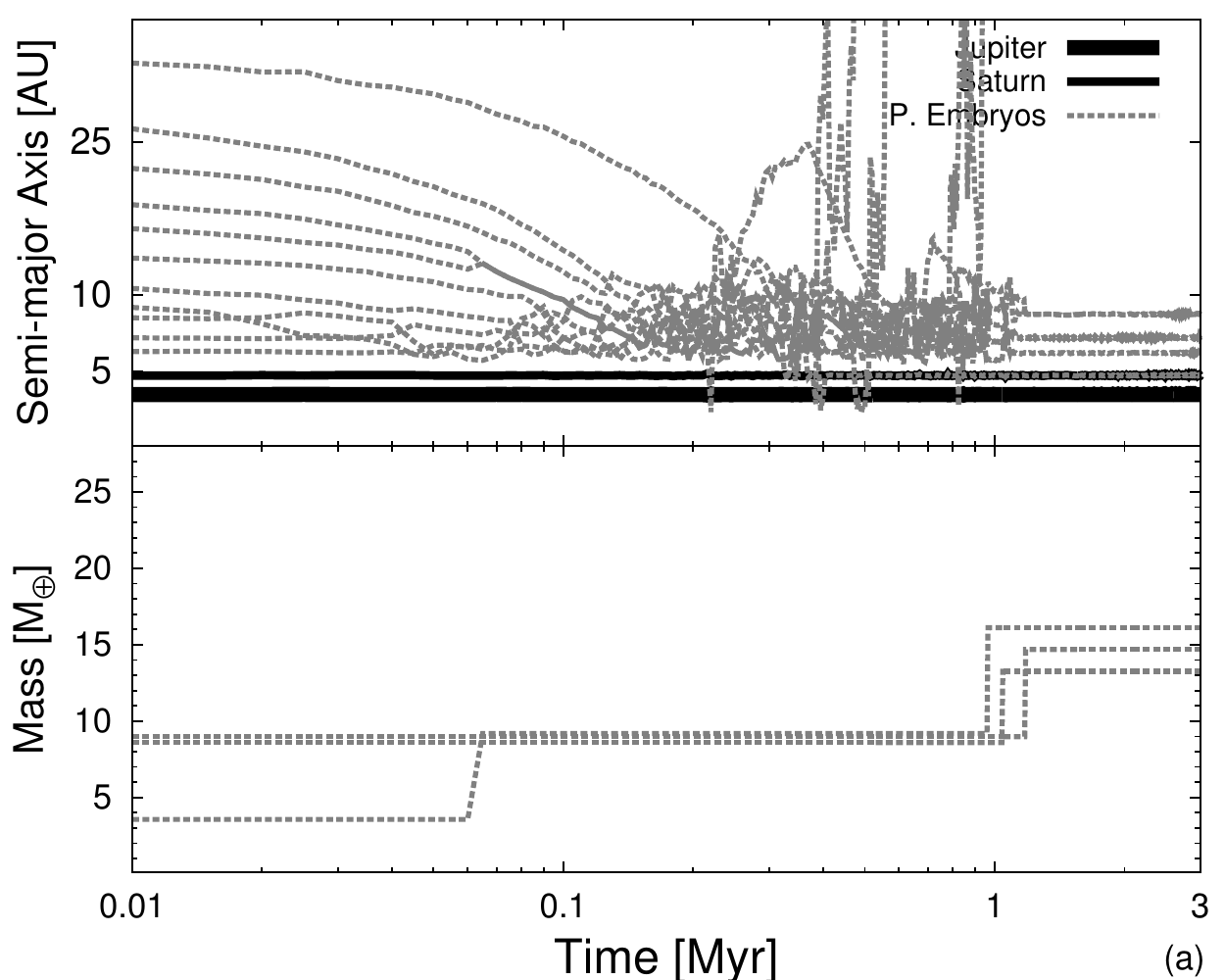}\hspace{0.8cm}
\includegraphics[scale=.6]{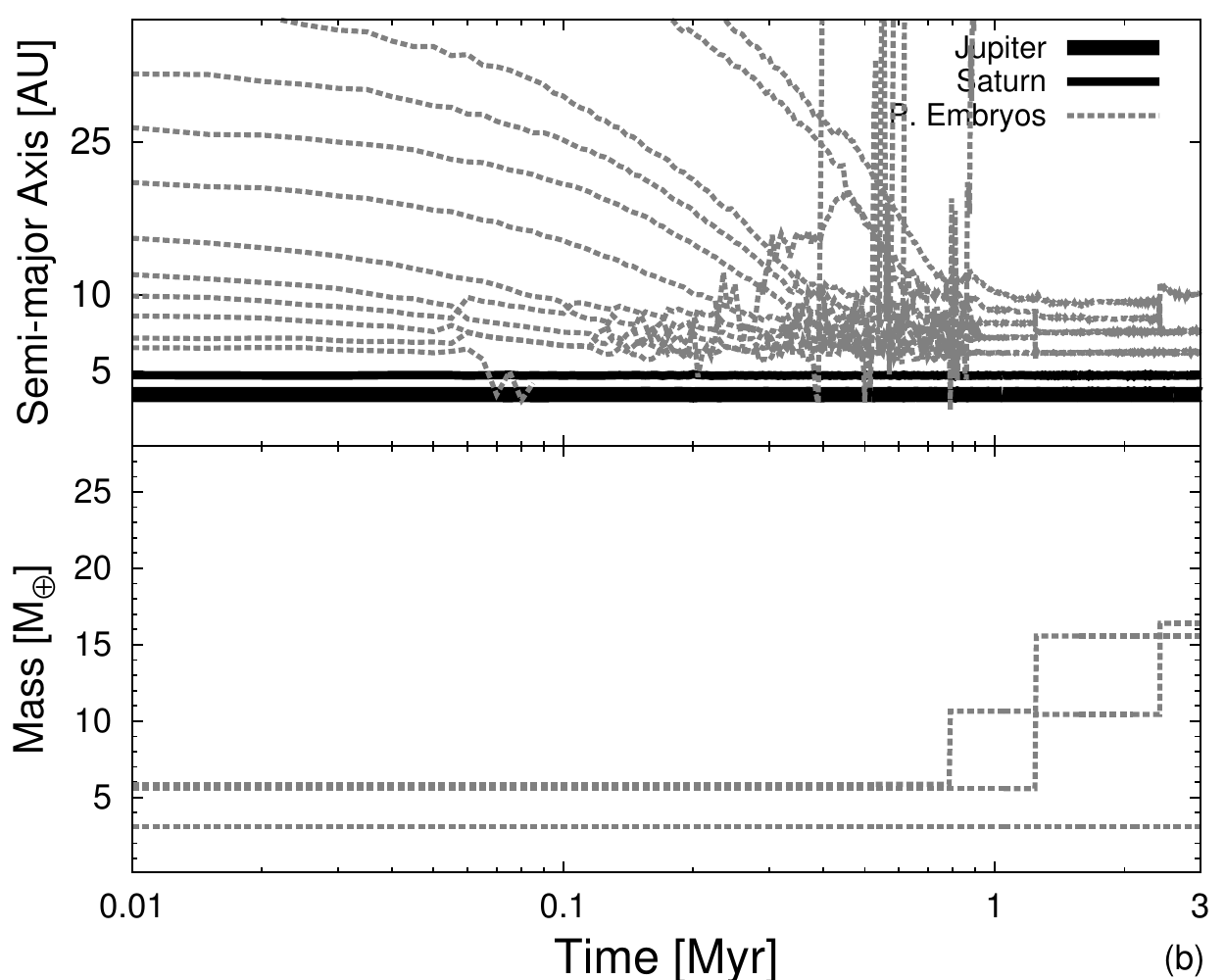}
\caption{Same that Figure 5 but in simulations where planetary embryos have initially different masses. In both simulations the gas dissipates in 3 Myr and the initial total mass carried by planetary embryos about 60 Earth masses.}
\end{figure}

Figure 10 shows that the  mean initial number of planetary embryos in our simulation of set V1 is about 10.5 planets  while in the set V2 this number is about 13. The horizontal error bars show the range over which the mean initial number of planetary embryos is calculated. In the vertical axis, Figure 10 also shows the mean final number of planetary cores  and also the range over which this value is calculated (vertical error bars).

Figure 11 shows the mean mass of the first innermost and second innermost planets formed beyond Saturn. The mass ratio between the mean mass of the first innermost core and second one beyond Saturn is 1.6 for V1 and 1.7 for V2. 
Comparing with our other results, this suggests that allowing a varied mass distribution may be almost equally good to simulations considering initially a population of planetary embryos with identical masses. In fact, the fraction  of good Uranus-Neptune analogs, as defined previously, produced in these simulations are  20\% (9\%) and 6\% (5\%) for set V1 and V2, respectively. As before, the values shown in brackets show the fraction of our simulations where at least one of the planetary cores experienced two collisions, both have masses larger or equal to 12${\rm M_{\oplus}}$ and the mass ratio between the two analogs is between 1 and 1.35 . However, we cannot fail noticing that simulations that successfully produced Uranus and Neptune analogs had initially massive planetary embryos in the system (eg. Figure 9), similar to the mass of planetary embryos in simulations starting with a single-mass component (${\rm \gtrsim 6M_{\oplus}}$).

\begin{figure}[H!]
\centering
\includegraphics[scale=.6]{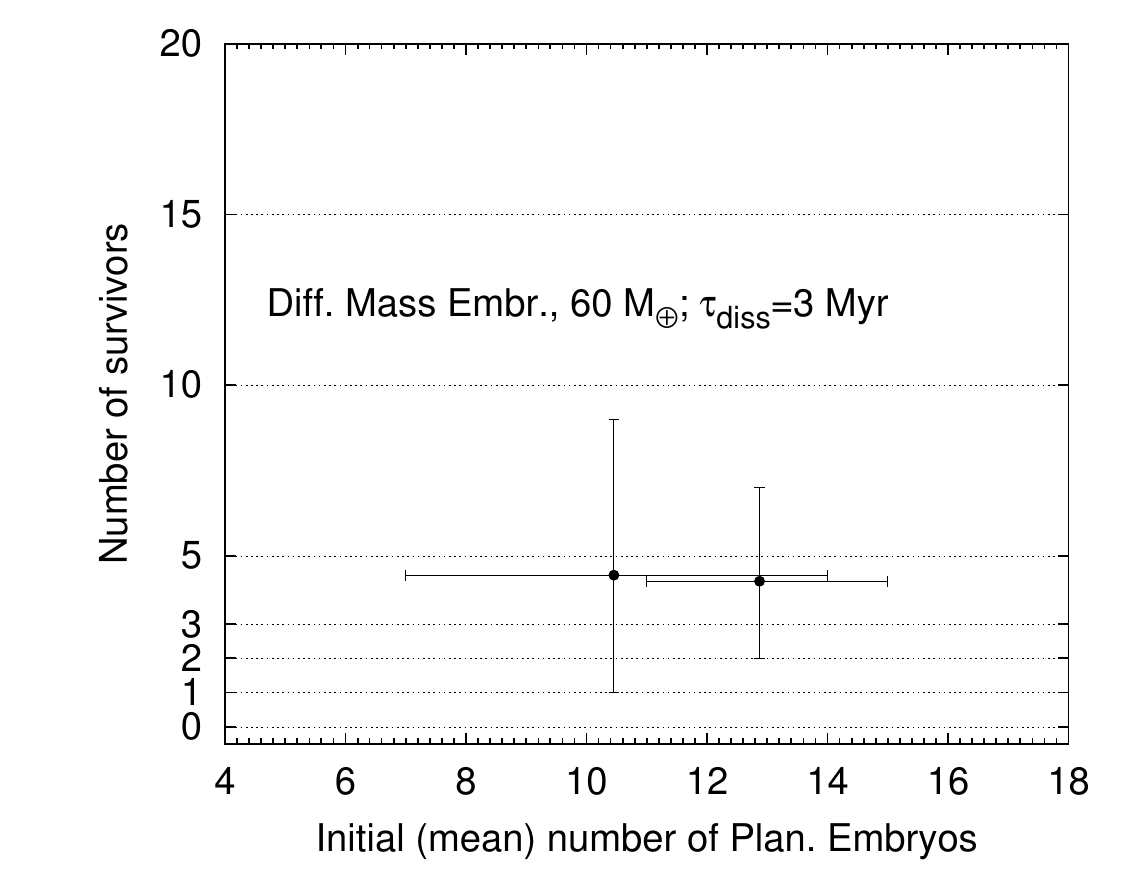}
\caption{Similar to Figure  3  but for simulations considering initially different mass for the planetary embryos. The horizontal axis show the initial (mean) number of planetary embryos.  The points show the the mean final number of planetary cores beyond Saturn for V1 and V2. The horizontal error bars show the range of values over which the mean initial number of planetary embryos is calculated. The vertical bars show the range of values over which the mean final number of planets is calculated.}
\end{figure}

\begin{figure}[H!]
\centering
\includegraphics[scale=.6]{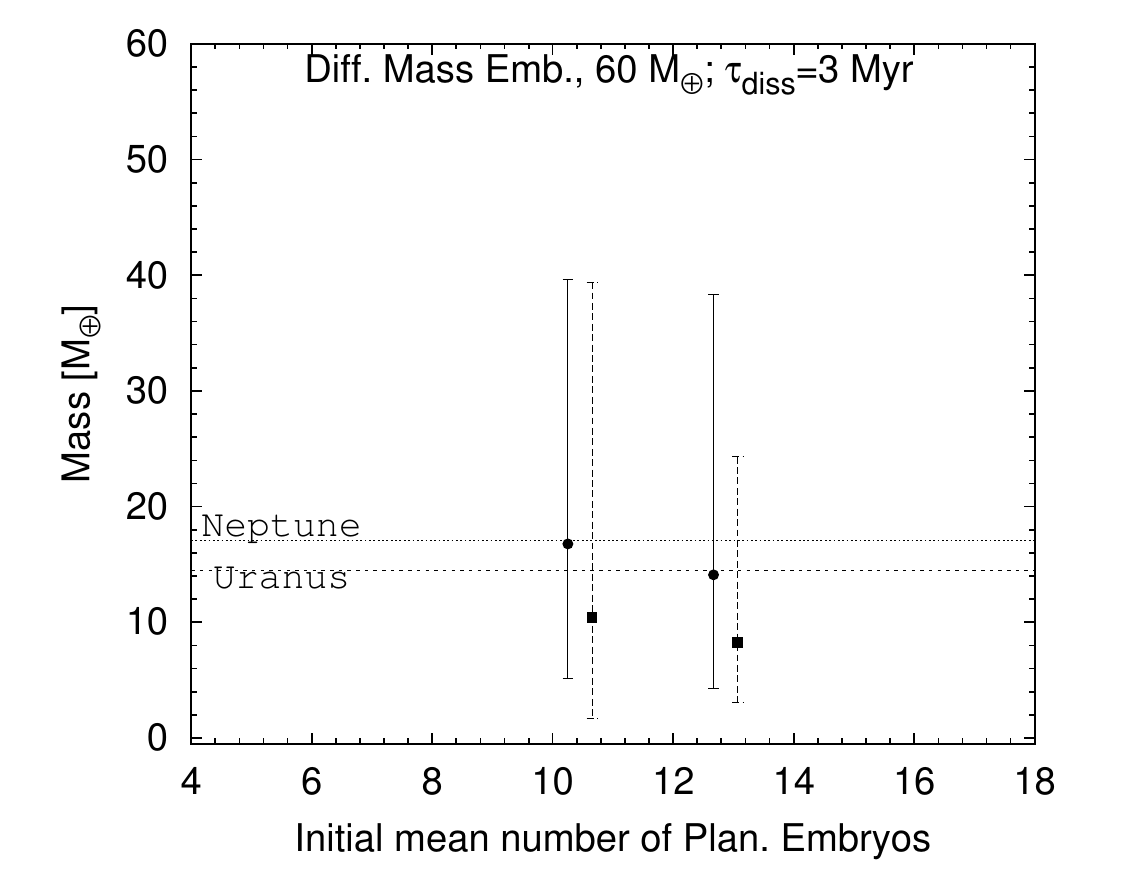}
\caption{Similar to Figure 4 but for simulations considering initially different mass for the planetary embryos. In this case we plot the initial mean number of planetary embryos (the range over which this number is calculated is shown by the horizontal error bars in Figure 10) against the  mean mass of the innermost and second innermost planetary cores formed beyond Saturn. The range over which this mean value is calculated is shown by the vertical errorbars.}
\end{figure}

\section{Discussion}

Our simulations confirm that producing planet analogs to Uranus and Neptune -- with large and comparable masses -- from a set of planetary embryos is indeed a difficult task. The major  challenge is not the individual masses of the simulated planets but their mass ratio. This is consistent with what Jakubík et al. (2012) found.

But unlike Jakubík et al. we also find that planetary embryos usually remain beyond the orbits of Jupiter and Saturn. The giant planets act as an efficient dynamical barrier (see Izidoro et al., 2015) that prevents embryos from jumping across their orbits.  The reason for this main difference with respect to the results of Jakubík et al. is our use of a more realistic surface density profile of the gaseous disk, as well as more realistic migration and damping forces. The low probability of penetration of an embryo into the inner solar system is a positive aspect of our results, because Izidoro et al. (2014) showed that the migration of a super-earth through 1 AU would have prevented the formation of the Earth, unless its migration occurred very rapidly.

 According with our scenario the success rate in producing Uranus and Neptune analogs varies significantly depending on the initial number of planetary embryos in the system, their initial and total masses. 
Our best results in terms of mass ratio were obtained in simulations considering initially planetary cores with masses equal or larger than  ${\rm 3~M_\oplus}$. In fact, ${\rm 6~M_\oplus}$ seems to be the best initial mass for coming close to the real masses and mass ratio of Uranus and Neptune. However, as observed for our simulations considering an initial distribution of planetary embryos with different masses an initial distribution of planetary embryos with different masses in a mass range between 3 and ${\rm 6~M_\oplus}$ or 1 and ${\rm 10~M_\oplus}$ may be similarly good.

The requirement that the initial embryos had a mass of the order of  ${\rm 5~M_\oplus}$ may shed doubts on the interest of our result. In fact, producing multiple ${\rm \sim 5~M_\oplus}$ embryos may be equally unlikely as forming directly two embryos with Uranus/Neptune masses.  This may not be possible by planetesimal accretion (Levison et al., 2010), but may be feasible by pebble accretion (Lambrechts and Johansen, 2012, 2014). We believe that the advantage of forming Uranus and Neptune from a set of smaller (although massive) embryos is that one can explain by giant impacts the origin of the large obliquities of Uranus and Neptune. In a significant fraction of our simulations (Table 1 and 2) the final planets indeed suffered at least one giant collision. 

It is quite interesting that the best scenario for the formation of Uranus and Neptune requires a population of planetary embryos of about ${\rm\sim 5~M_\oplus}$ (between 3 and ${\rm 6~M_\oplus}$). Recent studies (Youdin, 2011; Fressin et al., 2013; Petigura et al., 2013; Weiss \& Marcy, 2014; Silburt et al., 2014)  have shown that that the size distribution of extrasolar planets peaks at about $\sim$2 Earth radii (between 1.5 and $<$3.0). This same pattern is clearly present in the current planet candidate population (Burke et al., 2014). In fact, using the mean of the observed mass-density relation, a 2.0 Earth radii planet is equivalent to a $\sim$5 Earth mass planet (Weiss \& Marcy, 2014; Hasegawa \& Pudritz, 2014). Thus, it may be tempting to conjecture that ${\rm \sim 5 M_{\oplus}}$ is the typical mass of planetary embryos formed in the protoplanetary disk.

The typical dynamical evolution of our simulations  shows that a few embryos are scattered  beyond ~100 AU. In our simulations we remove these objects. In reality, if the solar system formed within a stellar cluster, with a significant probability (a few to 15\%) these planets could be decoupled  by stellar perturbations from Jupiter and Saturn and remain trapped on orbits with semi major axis of a few 100 to few 1000 AU (Brasser et al., 2006, 2012). Thus, if Uranus and Neptune were formed from a system of multi-Earth-mass planetary embryos our simulations may explain the existence today of a  primordial scattered planetary embryo on a distant orbit. The existence of such an object has been invoked to explain the observed orbital properties of the most distant trans-Neptunian objects  (eg. Gomes et al., 2006; Lykawka \& Mukai, 2008; Trujillo \& Sheppard, 2014). 

As in Jakubík et al. some of our simulations produced planetary cores in orbital resonance 1:1 with another planetary embryo. This was observed in about 5-20\% of our simulations that produced good Uranus and Neptune analogs. This  kind of orbital configuration is generally formed by the innermost planetary core beyond Saturn, which typically is the largest one, and a non-grown or partially grown planetary embryo (eg. Figure 5-f), but in a smaller fraction of cases we do observe the formation of a coorbital system with the second innermost planetary core beyond Saturn.  Such planetary arrangement is in contrast with the actual state of our solar system. However, as discussed in Jakubík et al., the smaller coorbital body would probably be removed during a later dynamical instability between the giant planets.

Most of our simulations that successfully produced Uranus-Neptune analogs formed more than 2 objects beyond Saturn (see for example Figure 5). This is because, as shown in Figure 3, simulations initially with 5 or 10 planetary embryos tend to end with 3 planetary objects beyond Saturn, on average.  In our model, the extra bodies are in general leftover planetary embryos that did not grow. This is partially consistent with models of the evolution of the solar system that consider the the solar system lost one or more ice giants (Nesvorny, 2011; Nesvorny \& Morbidelli, 2012). The main difference is that in our case the additional planets are in most cases original embryos, so they are smaller than Uranus and Neptune unlike what is considered in those works. Moreover, in our simulations the additional planets tend to be beyond Uranus and Neptune while in the best simulations of Nesvorny and Morbidelli (2012), they are placed in between.

One caveat of our results is  that about $\sim$35\% of the bodies that suffered at least one collision in our simulations have an obliquity either near zero (< 10 degrees) or 180 degrees (between 170 and 180 degrees). The remaining $\sim$65\% of our planetary cores, however, shows an isotropic obliquity distribution. This is a consequence of the  tidal inclination damping (Eq. 17) felt by the planetary embryos in our simulations. In our model, planetary embryos/cores may have their orbital inclinations damped to very low values which favors subsequent collisions with other objects  to happen near the equator (between two bodies with low orbital inclination ). The latter results in a small tilt to the final planet. Despite important, we do not consider that that issue invalidates our model. This drawback could be easily eliminated  if in reality the gaseous disk was turbulent (eg. Nelson, 2005). The intensity of this turbulence only needs to be strong enough to keep planetary embryos/cores around 10 AU with inclinations larger than $\sim$6e-4 degrees (see Figure 8). Thus, the needed stirring mechanism could be weak enough  not to affect the other properties of the dynamics, nor the process of pebble accretion, because the  velocity dispersion it would need to generate could be very small, about two orders of magnitude smaller than the deviation of the orbital velocity of the gas from the Keplerian velocity.

\section{Conclusions}

It remains a challenge to directly simulate the formation of Uranus and Neptune. Their growth by planetesimal accretion seems impossible both on their current orbits (Levison and Stewart, 2001) and on orbits at $\sim 10$ AU (Levison et al., 2010). The process of pebble accretion seems to be more efficient and is a promising mechanism for forming massive objects within the protoplanetary disk lifetime (Lambrechts and Johansen, 2012, 2014). However, it is unlikely that Uranus and Neptune formed purely by pebble accretion. Rather, the large obliquities of Uranus and Neptune suggest that both planets suffered giant impacts during their growth history (e.g. Slattery, 1992; Morbidelli et al., 2012).

In this paper we investigated the accretion of Uranus and Neptune by mutual collisions between multi-Earth-mass planetary embryos formed originally beyond the orbit of Saturn. These simulations correspond to the phase where the gaseous protoplanetary disk was still present but disappearing. Our simulations were performed using a N-body code adapted to take into account the effects of gas on the planetary embryos. Our protoplanetary disk is represented by a 1-D (radial) gas surface density profile, obtained by averaging over the azimuthal direction the result of a hydrodynamical simulation accounting from the presence of Jupiter and Saturn. The effects of type I migration, eccentricity and inclination damping on the orbits of the protoplanetary cores have been incorporated in our code in a way that had been previously calibrated to match the effects observed in hydrodynamical simulations and account for the shape of the density distribution of the gas-disk sculpted by the giant planets. We have performed simulations considering Jupiter and Saturn on non-migrating orbits and simulations considering these two giant planets migrating outwards (Walsh et al., 2011).

Our best results regarding the formation of analogs of Uranus and Neptune were obtained considering initially 5 or 10 planetary embryos with masses between 3 and  ${\rm 6~M_\oplus}$.  We tend to exclude the possibility of forming Uranus and Neptune from a system of more numerous (${\rm \sim20}$) but much smaller planetary embryos because  in all our simulation starting with 20 planetary embryos of ${\rm 3~M_\oplus}$ or smaller the final number of objects is on average larger than 5, and many cases as high as 10. In addition, the innermost planet formed in these simulations is usually very small.

With the exception of the simulations starting with 20 embryos  and a total mass in planetary embryos equal to ${\rm 30~M_{\oplus}}$, we produce in general at least one planet with a mass comparable to or larger than those of Uranus and Neptune. Most of our simulations do not show the scattering of an embryo into the inner solar system, which is consistent with the observed structure of the terrestrial planet system (Izidoro et al., 2014).  The most challenging property to match is the mass ratio between Uranus and Neptune.  In a significant fraction of the cases, however, we produce  two planets with a mass ratio between 1 and 1.5 (or even between 1 and 1.35), suggesting that the actual Uranus/Neptune mass configuration, although not typical, does have a significant probability to occur within this scenario.

\begin{acknowledgements}
We are very grateful to an anonymous referee for comments that helped us to substantially improve an earlier version of this paper. A.~I. thanks financial support from CAPES Foundation (Grant:~18489-12-5). A.~M., S.~R, F.~H and A.~P thank the Agence Nationale pour la Recherche for support via grant ANR-13-BS05-0003-01 (project MOJO). We also thank the CRIMSON team for managing the mesocentre SIGAMM of the OCA, where these simulations were performed. 
\end{acknowledgements}


\end{document}